\documentclass[12pt]{article}

\usepackage[ruled]{algorithm2e}
\usepackage{amsmath}
\usepackage{amssymb}
\usepackage{a4}
\usepackage{authblk}
\usepackage{amsfonts}
\usepackage{color}
\usepackage{booktabs}
\usepackage{caption}
\usepackage{graphicx}
\usepackage{hyperref}
\usepackage{mathrsfs}
\usepackage{multirow}
\usepackage{pdflscape}
\usepackage{subfigure}
\usepackage{theorem}
\usepackage[flushleft]{threeparttable}
\usepackage{url}
\usepackage{setspace}
\usepackage{geometry}
\newtheorem{theorem}{Theorem}
\newtheorem{lemma}[theorem]{Lemma}

\newtheorem{rem}[theorem]{Remark}
\newcommand{\comment}[1]{}

\newcommand{\ignorethis}[1]{}

\def \bfX {\mathbf{X}}
\def \bfx {\mathbf{x}}

\def \bfy {\mathbf{y}}

\def \bfs {\mathbf{s}}

\def \bflambda {\boldsymbol{\lambda}}

\def \bfSigma {\boldsymbol{\Sigma}}

\def \bfbeta {\boldsymbol{\beta}}

\def \bfepsilon {\boldsymbol{\epsilon}}

\def \bfR {\mathbf{R}}

\def \bfS {\mathbf{S}}
\def \bfv {\mathbf{v}}

\def \bfD {\mathbf{D}}

\def \bfI {\mathbf{I}}

\def\bfv{\mathbf{v}}

\date{}

\begin{document}
\setlength{\textheight}{575pt}
\setlength{\baselineskip}{23pt}


\title{REMI: Regression with marginal information and its application in genome-wide association studies}


\author[1]{Jian Huang}
\affil[1]{Department of Applied Mathematics, Hong Kong Polytechnics University}
\author[2]{Yuling Jiao}
\affil[2]{Department of Statistics, Zhongnan University of Economics and Law}
\author[3]{Jin Liu}
\affil[3]{Centre for Quantitative Medicine, Duke-NUS Medical School}
\author[4]{Can Yang}
\affil[4]{Department of Mathematics, Hong Kong University of Science and Technology}

\maketitle

\begin{abstract}
In this study, we consider the problem of variable selection and estimation in high-dimensional linear regression models when the complete data are not accessible, but only certain marginal information or summary statistics are available. This problem is motivated from the Genome-wide association studies (GWAS) that have been widely used to identify risk variants underlying complex human traits/diseases. With a large number of completed GWAS, statistical methods using summary statistics become more and more important because of restricted accessibility to individual-level data sets. Theoretically guaranteed methods are highly demanding to advance the statistical inference with a large amount of available marginal information.  Here we propose an $\ell_1$ penalized approach, REMI, to estimate high dimensional regression coefficients with marginal information and external reference samples. We establish an upper bound on the error of the REMI estimator, which has the same order as that of the minimax error bound of Lasso with complete individual-level data. In particular, when marginal information is obtained from a large number of samples together with a small number of reference samples, REMI yields good estimation and prediction results, and outperforms the Lasso because the sample size of accessible individual-level data can be limited. Through simulation studies and real data analysis of the NFBC1966 GWAS data set, we demonstrate that REMI can be widely applicable. 
The developed R package and the codes to reproduce all the results are available at \href{https://github.com/gordonliu810822/REMI}{https://github.com/gordonliu810822/REMI}

\end{abstract}
Keywords: Genome-wide association studies, marginal information, high dimensional regression.

\newpage

\section{Introduction}

High dimensional regression has been widely applied in various fields, such as medicine, biology, finance and marketing \cite{Hastie2009elements}. Consider the linear regression model that relates a response variable $Y$ to a vector of $p$ predictors $X=(X_1,\dots,X_p)^T$:
\begin{equation}
\label{Reg1}
Y = \sum^p_{j=1} X_j \beta^*_j + \epsilon,
\end{equation}
where $\bfbeta^* = (\beta^*_1,\dots,\beta^*_p)^T$ is the vector of regression coefficients and $\epsilon$ is the random error term with mean zero and noise level  $\sigma_{\epsilon}^2.$ In most applications, the data set is comprised of an $n\times p$ matrix $\bfX$ for variables in $X$ and a vector $\bfy=(y_1,\dots,y_n)^T$ for response $Y$  collected from $n$ individuals. Given the individual-level data $\{\bfX,\bfy\}$, there exist  convex  \cite{Tibshirani:1996,CandesTao:2007} and nonconvex  \cite{FanLi:2001,Zhang:2010a} penalized methods for estimating $\bfbeta^*$  with theoretical guarantee  \cite{ZhaoYu:2006,MeinshausenBuhlmann:2006,ZhangHuang:2008,BickelRitovTsybakov:2009,ZhangZhang:2012},
just name a few.  Also see the monographs \cite{buhlmann2011statistics,hastie2015statistical} and the references therein.

Motivated from the applications in human genetics, 
we consider the problem of estimating $\bfbeta^*$ when the individual-level data $\{\bfX,\bfy\}$ is not accessible but the marginal information is available, such as $\bfX_j^T \bfy$ and $\bfX^T_j \bfX_j$, $j = 1,\dots,p$, where $\bfX_j$ is the $j$-th column of $\bfX$. For this reason, we refer to our problem formulation as  "\underline{re}gression with \underline{m}arginal \underline{i}nformation" (REMI). To make our formulation feasible, we also assume that information of the covariance structure of variables in $X$ can be estimated via a reference  panel data set in the form of  an $n_{\textrm{r}}\times p$ data matrix $\bfX_{\textrm{r}}$, where $n_{\textrm{r}}$ is the number of samples from the reference panel and $n_{\textrm{r}}\ll p$. A natural question arises: Without accessing the individual-level data, can we use marginal information together with the reference data $\bfX_{\textrm{r}}$ to estimate $\bfbeta^*$, assuming observations in $\bfX_{\textrm{r}}$ and $\bfX$ are from the same distribution?

In particular, our problem arises 
in genome-wide association studies (GWAS), which
 have been conducted over the past decade to study the genetic basis of human complex phenotypes, including both quantitative traits and complex diseases \cite{hindorff2009potential,welter2014nhgri,visscher201710}. As of April, 2018, more than 59,000 unique phenotype-variant (typically Single Nucleotide Polymorphism, or SNP in short) associations have been reported in about 3,300 publications of GWAS (see the GWAS Catalog database \url{https://www.ebi.ac.uk/gwas/}). 
An important lesson from GWAS \cite{yang2010common,visscher2012five,visscher201710} is that complex phenotypes are highly polygenic, that is, they are often affected by many genetic variants with small effects. Well-known examples include human height \cite{wood2014defining}, psychiatric disorders \cite{gratten2014large}, as well as diabetes \cite{fuchsberger2016genetic}. Due to the polygenicity, variants with small effects largely remain undiscovered yet and large sample sizes are required in exploring genetic architectures of complex phenotypes. Researchers world-wide are forming large genomic consortia, such as the Genetic Investigation of ANthropometric Traits (GIANT) Consortium and the psychiatric genomic consortium (PGC), to maximize sample size, aiming at a deeper understanding of
the genetic architecture of complex phenotypes.

Although much efforts have been made for data sharing, it is still very difficult for a research group to fully access the individual-level genotype data available in a consortium. For example, a core research group from the GAINT consortium reported that they can only access genotype data from about 44,000 individuals \cite{yang2015genome} while the total sample size is more than 250,000 for the consortium \cite{wood2014defining}. There are several reasons for the restricted access to the individual-level data. First, privacy protection is always a big concern in sharing individual-level genotype data. Second, it is often time-consuming to achieve an agreement on data-sharing among different research groups. Third, many practical issues arise in data transportation and storage. In contrast, summary statistics from GWAS are widely available through many public gateways \cite{editor2012ng}, e.g., the download session at the GWAS Catalog \url{https://www.ebi.ac.uk/gwas/downloads/summary-statistics}. Because these summary statistics (e.g., estimated effect sizes, standard errors, and $z$-values) are often generated by simple linear regression analysis, summary statistics are essentially marginal information. 

To meet the great demand of data analysis in GWAS, various statistical methods have been proposed to utilize marginal information. Using a few hundreds of human genome data from the 1000 Genome Project as a reference panel, information on the correlation structure of genetic variants (typically using ``linkage disequilibrium'' in genetics, or LD for short) becomes available. This allows these methods to bypass the individual-level data but only use marginal information. Here we roughly divide these methods into three categories:  (a)  Methods for heritability estimation. Heritability of a phenotype quantifies
the relative importance of genetics and environment to the phenotype \cite{visscher2008heritability}. When individual-level data is accessible, linear mixed model (LMM)-based approaches (e.g., GCTA \cite{yang2010common,yang2011gcta}) are widely used for heritability estimation \cite{lee2011estimating}. In the absence of individual-level data, Bulik et al. \cite{bulik2015ld} first introduced the LD score regression, named LDScore, for heritability estimation only using summary statistics and the reference data from the 1000 Genome Project.
Based on the minimal norm quadratic unbiased estimation criteria, Zhou \cite{zhou2016unified} proposed a novel method of moments, MQS, for variance component estimation with summary statistics. (b) Methods for association mapping. Heritability estimation provides a global measure which quantifies the overall contribution from genetic factors while association mapping is to localize genetic variants associated with a given phenotype. Recently, a few statistical methods have been proposed for association mapping based on summary statistics, including FGWAS \cite{pickrell2014joint}, PAINTOR \cite{kichaev2014integrating}, CAVIAR \cite{hormozdiari2014identifying}, and CAVIARBF \cite{chen2015fine}. Although these methods are very useful for performing association mapping on summary statistics, they still have their limitations. On one hand, they adopted some ad-hoc ways to reduce computational cost. For example, to avoid a combinatorial search, FGWAS assumes that there is only one causal signal in an LD block and PAINTOR searches no more than two causal variants in its default setting. On the other hand, statistical analysis is oversimplified to overcome estimation difficulties. For example, the non-centrality parameter in PAINTOR  and the variance components in CAVIAR and CAVIARBF are pre-fixed rather than adaptively estimated from data. (c) Methods for effect size estimation and risk prediction. Recently, Vilhj{\'a}lmsson et al. \cite{vilhjalmsson2015modeling} proposed a Bayesian method, LDpred, for effect size estimation and risk prediction by accounting for LD. Along this line, Hu et al. \cite{hu2017leveraging} further introduced AnnoPred to improve LDpred by incorporating functional information in human genome. However, neither LDpred or AnnonPred should be considered as a marginal-information-based method because it requires individual-level data as validation data for its parameter tuning.

Although the existing statistical methods have shown a good empirical performance in GWAS data analysis, there are a number of open questions on REMI. First, the sample size of the reference panel is often very small. For example, there are only about 370 samples from the 1000 Genome Project that can be used as reference for analyzing GWAS data in European ancestry. It remains unclear why such a small sample size is often good enough for exploring the correlation structure of a large number of variables (i.e., SNPs). Second, the theoretical properties of the existing methods for effect size estimation and prediction error are not clear. Third, the sampling-based algorithms are often time-consuming as they need to run thousands of Markov Chain Monte Carlo (MCMC) iterations \cite{zhu2016bayesian}. In this paper, we propose a unified framework to address the above open questions. The rest of this paper is organized as follows: In Section~\ref{model}, we introduce our REMI model and discuss REMI in GWAS. In Section~\ref{empirical}, we present an efficient coordinate descent algorithm following by discussion on some practical issues. In Section~\ref{theory}, we establish the error bound and prediction error of the proposed method. In particular, our theoretical results explain why a small number of samples (i.e., $n_{{\textrm{r}}}$) from the reference panel can be good enough for effect size estimation and risk prediction.  In Section~\ref{emp_results}, we show the results from both simulation studies and real data analysis.

\section{The REMI model}
\label{model}
\subsection{The REMI model}

For the linear regression model \eqref{Reg1},
if the individual-level data $(\bfy, \bfX)$ is available,
a basic approach for estimating $\bfbeta^*$ in high-dimensional settings is the Lasso
\cite{Tibshirani:1996}.
The Lasso estimator is given by
\begin{align}\label{regularized_reg}
\widehat{\bfbeta} = \arg\min_{\bfbeta} \frac{1}{n} \|\bfy - \bfX \bfbeta\|^2 + \lambda \|\bfbeta\|_1, 
\end{align}
where $\|\cdot\|_1$ is the $\ell_1$ norm and $\lambda \ge 0 $ is a regularization parameter.
In our problem, however, the individual-level data $\{\bfX, \bfy\}$ is not accessible.
Hence, direct application of the Lasso is not feasible here.
We note that several other important penalized methods have been proposed, including
 SCAD \cite{FanLi:2001} and MCP \cite{Zhang:2010a}. We will focus on the Lasso penalty below, although our proposed approach can also be based on the other penalties.

We now describe our proposed REMI model with the Lasso penalty.
Rewrite (\ref{regularized_reg}) as
\begin{align}\label{regularized_reg2}
\widehat{\bfbeta} &= \arg\min_{\bfbeta} \frac{1}{n} (\bfbeta^T \bfX^T \bfX \bfbeta -2\bfbeta^T \bfX^T \bfy + \bfy^T\bfy) + \lambda \|\bfbeta\|_1 \nonumber\\
& = \arg\min_{\bfbeta} \bfbeta^T \bfX^T \bfX \bfbeta/n -2\bfbeta^T \bfX^T \bfy/n + \lambda \|\bfbeta\|_1,
\end{align}
where the second term only involves the inner product of the optimization variable $\bfbeta$ and marginal information, say, $ \widetilde{\bfy} = \bfX^T \bfy/n$, which we assume is available. The difficulty comes from the first term, where $\bfX^T\bfX/n$ is unknown since $\bfX$ is not observed. Motivated by the application in GWAS, we assume that there exists a reference $n_{\textrm{r}} \times p$ data matrix $\bfX_{\textrm{r}}$, where the rows of $\bfX_{\textrm{r}}$ are i.i.d. and have the same distribution with covariance matrix $\bfSigma$ as the rows of $\bfX$.
Therefore, both $\widehat\bfSigma = \bfX^T\bfX/n$ and $\widehat{\bfSigma}_{\textrm{r}}  = \bfX_{\textrm{r}}^T\bfX_{\textrm{r}}/n_{\textrm{r}}$ can be viewed as estimators of $\bfSigma$.
So we propose to solve the following optimization problem to estimate $\bfbeta^*$:
\begin{align}\label{remi_1}
\widehat{\bfbeta}^{\textrm{c}} = \arg\min_{\bfbeta} \bfbeta^T \bfX_{\textrm{r}}^T \bfX_{\textrm{r}} \bfbeta/n_{\textrm{r}} -2\bfbeta^T\widetilde{\bfy}  + \lambda \|\bfbeta\|_1,
\end{align}
where $\widehat{\bfbeta}^{\textrm{c}}$ denotes the estimator using the reference covariance matrix. Clearly, the above model (\ref{remi_1}) only uses the marginal correlation between $\bfX$ and $\bfy$, with the covariance matrix estimated by an external reference panel $\bfX_{\textrm{r}}$.

\subsection{REMI in GWAS}\label{remi-gwas}
In the context of GWAS, the available marginal information may not be $\widetilde{\bfy} = \bfX^T \bfy/n$ but summary statistics $\{\widehat{\beta}^{\textrm{m}}_{j}, \hat{s}^2_j\}_{j=1,\dots,p}$ from univariate linear regression:
\begin{align*}
 \widehat{\beta}^{\textrm{m}}_{j} = (\bfX^T_j\bfX_j)^{-1}\bfX^T_j \bfy, \quad \hat{s}^2_j = (n\bfX^T_j\bfX_j)^{-1}(\bfy-\bfX_j \widehat{\beta}^{\textrm{m}}_{j})^T(\bfy-\bfX_j\widehat{\beta}^{\textrm{m}}_{j}),
\end{align*}
where superscript $^{\textrm{m}}$ is used to denote marginal information.
$\widehat{\beta}_{j}^{\textrm{m}}$ and $\hat{s}^2_j$ are the estimated effect size and its variance for SNP $j$, respectively. Due to 
the polygenicity of many complex phenotypes, the standard errors  can be well approximated by $\hat{s}_j \approx \sqrt{(n\bfX^T_j\bfX_j)^{-1}\bfy^T\bfy}$ (Zhu and Stephens 2016). Let $\widehat{\bfbeta}^{\textrm{m}} = [\widehat{\beta}_{1}^{\textrm{m}}, \dots, \widehat{\beta}_{p}^{\textrm{m}}]^T$, $\hat{\bfs}^2 = [\hat{s}^2_1,...,\hat{s}^2_p]^T$ be   the vectors collecting estimated effect sizes and estimated variance, respectively,
and $\widehat\bfS$ be a $p\times p$ diagonal matrix with $\hat{s}_j$ being its $j$-th diagonal element. Further, we introduce a $p\times p$ diagonal matrix $\widehat{\bfD}=\mathrm{diag}(\hat{d}_j)$ with its $j$-th diagonal element being the sample standard deviation of $\bfX_j$, i.e., $\hat{d}_j = \sqrt{\frac{\bfX_j^T \bfX_j}{n}}$, and correlation matrix $\widehat{\bfR}=[\hat{r}_{jk}]\in\mathbb{R}^{p\times p}$ with $\hat{r}_{jk}=\frac{\bfX_{j}^T \bfX_{k}}{(\bfX_{j}^T \bfX_{j})^{1/2}(\bfX_{k}^T \bfX_{k})^{1/2}}$. Noticing that $\hat{d}_j^{2} \widehat{\beta}_{j}^{\textrm{m}} = \bfX^T_j\bfy/n$ and $n^2 \hat{d}^2_j \hat{s}^2_j \approx \bfy^T\bfy$, the REMI formulation (\ref{regularized_reg2}) becomes
\begin{equation}
\begin{aligned}
\widehat{\bfbeta} &= \arg\min_{\bfbeta}\, \bfbeta^T \bfX^T \bfX \bfbeta/n -2\bfbeta^T \bfX^T \bfy/n + \lambda \|\bfbeta\|_1, \nonumber\\
&= \arg\min_{\bfbeta}\, \bfbeta^T \widehat{\bfD}\widehat{\bfR}\widehat{\bfD} \bfbeta -2\bfbeta^T \widehat{\bfD}^{2}\widehat{\bfbeta}^{\textrm{m}} + \lambda \|\bfbeta\|_1,\nonumber \\
&\approx\arg\min_{\bfbeta}\, \frac{\bfy^T\bfy}{n^2}\bfbeta^T \widehat{\bfS}^{-1}\widehat{\bfR}\widehat{\bfS}^{-1} \bfbeta -2 \frac{\bfy^T\bfy}{n^2}\bfbeta^T \widehat{\bfS}^{-2}\widehat{\bfbeta}^{\textrm{m}} + \lambda \|\bfbeta\|_1, \\
&=\arg\min_{\bfbeta}\, \bfbeta^T \widehat{\bfS}^{-1}\widehat{\bfR}\widehat{\bfS}^{-1} \bfbeta -2 \bfbeta^T \widehat{\bfS}^{-2}\widehat{\bfbeta}^{\textrm{m}} + \widetilde{\lambda} \|\bfbeta\|_1, \nonumber \\
\end{aligned}
\end{equation}
where $\widetilde{\lambda} = \frac{n^2}{\bfy^T\bfy} \lambda$ and the approximation holds in the case of polygenicity. Because $\widetilde\lambda$ is a tuning parameter that scales $\lambda$ with a constant factor ($\frac{n^2}{\bfy^T\bfy}$), we slightly abuse $\lambda$ for $\widetilde{\lambda}$ and propose to solve the following optimization problem
\begin{align}\label{remi_2}
\widehat{\bfbeta}^{\textrm{r}} = \arg\min_{\bfbeta}\, L(\bfbeta)  +\lambda \|\bfbeta\|_1, 
\end{align}
where $L(\bfbeta)=\bfbeta^T \widehat{\bfS}^{-1}\widehat{\bfR}\widehat{\bfS}^{-1} \bfbeta -2 \bfbeta^T \widehat{\bfS}^{-2}\widehat{\bfbeta}^{\textrm{m}}$,
and $\widehat{\bfbeta}^{\textrm{r}}$ denotes the estimates using correlation information.
Similar to REMI (\ref{remi_1}) in which covariance matrix $\widehat\bfSigma=\bfX^T\bfX/n$ needs to be estimated, here correlation matrix $\widehat{\bfR}$ needs to be estimated by samples from the reference panel $\bfX_{\textrm{r}}$.
We refer (\ref{remi_1}) as  REMI-C and (\ref{remi_2}) as the REMI-R, respectively.
\section{Algorithm and Practical Issues}
\label{empirical}
\subsection{Algorithm}

Here we adopt the widely used coordinate descent algorithm, which updates one parameter at a time, say $\widehat{\beta}^{\textrm{c}}_{j}$, keeping all other parameters fixed at their current values.
Thus the sub-problem for parameter $\widehat{\beta}^{\textrm{c}}_{j}$ can be written as
\begin{align}\label{remi_1_j}
\widehat{\beta}^{\textrm{c}}_{j}(\lambda)  = \arg\min_{{\beta}_{j}} \widehat{\sigma}_{jj} \beta_{j}^2  - 2\left(\widetilde{y}_j- \sum_{k\ne j} \widehat{\beta}_{k}^{\textrm{c}}{\widehat\sigma}_{jk}  \right){\beta}_{j} + \lambda|\beta_{j}|,
\end{align}
where $\widehat{\sigma}_{jk}$ is an element in  $\widehat{\bfSigma}_{\textrm{r}} = [\widehat{\sigma}_{jk}]\in\mathbb{R}^{p\times p}$.
 An efficient path algorithm can be developed based on the warm start and some other tricks as described in~\cite{friedman2010regularization}. In particular, we generate a sequence of $\bflambda = (\lambda_1,\dots,\lambda_D)$ equally spaced in logarithm with $\lambda_1 = \lambda_{\mathrm{max}}$ and $\lambda_D = \tau\lambda_{\mathrm{max}}$, where $\lambda_{\mathrm{max}}$ is the minimum $\lambda$ that shrinks all parameters to zero and $\tau$ is usually set to 0.05. For each $\lambda$, we use the solution of (\ref{remi_1}) from the last $\lambda$ value as warm start. The path algorithm is described in Algorithm \ref{alg2}.


\begin{algorithm}[H]
	\label{alg2}\caption{Path algorithm to solve REMI-C ~(\ref{remi_1}) with a sequence of $\bflambda = (\lambda_1,\dots,\lambda_D)$}
	\emph{Output}: Solution path for $\widehat{\bfbeta}^{\textrm{c}}(\bflambda)$.\\ 
	\For{$l=1,2,\dots,D$}{
		Initialize  $\widehat{\bfbeta}^{\textrm{c}}(\lambda_l)= \widehat{\bfbeta}^{\textrm{c}}(\lambda_{l-1})$, if $l > 1$; $\widehat{\bfbeta}(\lambda_l) = \boldsymbol{0}$, if $l = 1$
		
		\Repeat{Convergence}{
			\For{$j=1,2,\dots,p$}{
				\qquad  $\eta_j = \widetilde{y}_j- \sum_{k\ne j} \widehat{\beta}_{k}^{\textrm{c}}(\lambda){\widehat\sigma}_{jk}$ \\
			\qquad  $\widehat{\beta}_{j}^{\textrm{c}}(\lambda) \leftarrow \mathrm{S}(\eta_j,\lambda/2)/ {\widehat\sigma}_{jj}$ \\
	}}}
\end{algorithm}



Similar to REMI-C, 
an efficient coordinate descent algorithm can developed for solving REMI-R (\ref{remi_2}).
The efficient path algorithm is given in Algorithm \ref{alg4}.


\begin{algorithm}[H]
\label{alg4}\caption{Path algorithm to solve REMI-R  (\ref{remi_2}) with a sequence of $\bflambda = (\lambda_1,\dots,\lambda_D)$}
\emph{Output}: Solution path $\widehat{\bfbeta}^{\textrm{r}} (\bflambda)$.\\
\For{$l=1,2,\dots,D$}{
Initialize  $\widehat{\bfbeta}^{\textrm{r}}(\lambda_l) = \widehat{\bfbeta}^{\textrm{r}}(\lambda_{l-1})$, if $l > 1$; $\widehat{\bfbeta}^{\textrm{r}}(\lambda_l) = \boldsymbol{0}$, if $l = 1$

\Repeat{Convergence}{
\For{$j=1,2,\dots,p$}{
\qquad  $\eta_j = \frac{\widehat{\beta_{j}}^{\textrm{m}}}{\hat{s}_j^2}-\frac{1}{\hat{s}_j}\sum_{k\ne j} \frac{\widehat{\beta_{k}}^{\textrm{r}}\hat{r}_{jk}}{\hat{s}_{k}}$ \\
\qquad  $\widehat{\beta_{j}}^{\textrm{r}}(\lambda) \leftarrow \mathrm{S}(\eta_j,\lambda/2)\times \hat{s}_j^2$ \\
}}}
\end{algorithm}

\subsection{Reference panel}\label{refpanel}

In REMI-R model (\ref{remi_2}), it involves the cohort-based estimated correlation matrix. Based on the
nature of the correlation patterns of the SNPs,
$\bfR$ can be approximated by a block diagonal matrix. Specifically, we first partition the whole genome into
$L$ blocks ($L=1,703$ for European ancestry and $L=1,445$ for Asian ancestry, respectively \cite{berisa2016approximately}).
Then we calculate empirical correlation matrix $\widehat{\bfR}_{\mathrm{emp}}$ for each LD-block. To ensure a stable numerical result, we apply a simple shrinkage estimator to obtain $\widehat{\bfR}^{\mathrm{r}} = \kappa \widehat{\bfR}_{\mathrm{emp}} + (1-\kappa) \bfI$ within each block \cite{schafer2005shrinkage}, where we used $\kappa=0.9$ as default (the estimate of $\bfbeta^*$ is insensitive to $\kappa$ \cite{pasaniuc2014fast}). Thus, similar to \cite{zhu2016bayesian}, REMIs and its individual-level-data counterpart will produce approximately the same inferential results within a region. After plugging $\widehat\bfS$ and $\widehat\bfR^{\mathrm{r}}$ in (\ref{remi_2}), we can use the coordinate-descent algorithm to obtain $\widehat{\bfbeta}^{\mathrm{r}}(\lambda)$ (Algorithm \ref{alg4}).

\subsection{Choice of regularization parameter $\lambda$}
The REMIs have one regularization parameter $\lambda$. Here we briefly show how to choose this parameter for REMI-R and it is straightforward to develop the same strategy for REMI-C. Similarly to the Lasso solver \cite{friedman2010regularization}, we generate a sequence of $\bflambda$ from $\lambda_{\mathrm{max}}$ to $\tau\lambda_{\mathrm{max}}$, where $\lambda_{\mathrm{max}}$ is the minimum value of $\lambda$ that shrinks all parameters to zero and $\tau$ is  pre-specified with the default value at 0.05. Note that $\lambda_{\mathrm{max}} = \max \big\{2\hat\beta_j^{\mathrm{m}}/\hat{s}^2_j\big\}_{j=1,\dots,p}$. We search for optimal value $\lambda$ value using BIC,
\begin{equation}
  \mathrm{BIC}(\lambda_l) =  L(\widehat\bfbeta^{\mathrm{r}}(\lambda_l)) + \mathrm{log}(n)\mathrm{df}(\lambda_l).
\end{equation}
Zou et al. \cite{zou2007degrees} showed that the number of nonzero coefficients is an unbiased estimate for the degrees of freedom of Lasso. We choose $\mathrm{df}(\lambda_l)$ to be the number of nonzero coefficients given $\lambda_l$. To make fair comparison of REMIs with Lasso, we also use BIC to select the regularization parameter when individual-level-data is accessible.

\section{Theoretical properties}
\label{theory}
In this section, we give nonasymptotic bounds on the estimation error  $\|\bfbeta^{*}_{\mathcal{A}}-\widehat{\bfbeta}^{\textrm{c}}\|$ and the prediction error
$\| \bfX_\mathrm{new} \bfbeta^{*}_{\mathcal{A}}-\bfX_\mathrm{new} \widehat{\bfbeta}^{\textrm{c}}\|^2/n_{\mathrm{new}}$, where $ \mathcal{A} $ denotes index of significant entries of $\bfbeta^*$ and $\bfbeta^*_{\mathcal{A}}$ denote the vector supported on $\mathcal{A}$.

 Since in  real applications of genetic data the underlying signal is not exactly sparse but with many small components.  Here, we assume that the target $\bfbeta^*$ is weak sparse, i.e., in addition to some significant components indexed by $ \mathcal{A} $,  there  may be  many nonzero entries in $\bfbeta^*$  with very small magnitude, as indexed by $\mathcal{I} = \mathcal{A}^c$.
Let $\bfbeta^*_{\mathcal{I}}$ be the  vector supported on $ \mathcal{I}$. It is reasonable to assume that
\begin{equation}\label{weaksparse}
 s = |\mathcal{A}| \leq n, \quad  \|\bfbeta^*_{\mathcal{I}}\|_{\infty} \leq 2\sigma_{\epsilon}\sqrt{\frac{\log p}{n}},
\end{equation}
since the  signal whose magnitude  smaller than this order  is undetectable.
 Then
 \begin{equation}\label{dec}
 \widetilde{\bfy} = \widehat{\bfSigma} \bfbeta^* + \widetilde{\bfepsilon} = \widehat{\bfSigma} \bfbeta^*_{\mathcal{A}} + \widehat{\bfSigma} \bfbeta^*_{\mathcal{I}}+ \widetilde{\bfepsilon},
 \end{equation}
 where, $\widetilde{\bfepsilon} = \bfX^T \bfepsilon/n.$
Let $C_1 \geq  \|\textrm{diag}(\bfSigma)\|_{\infty}$, $C_2 \geq \max_{j=1,...p} \{\|\bfSigma_j\|_1\}$,  $C_3\geq \|\bfbeta^*_{\mathcal{A}}\|_1,$ $C_4 \geq \|\bfbeta^*_{\mathcal{I}}\|_1.$
Recall the restrict eigenvalue  \cite{Bickel:2009} of  $\widehat{\bfSigma}_{\textrm{r}}$ is defined as  $$\phi_{\widehat{\bfSigma}_{\textrm{r}}} = \min_{0\neq \bfv \in \mathcal{C}_{\mathcal{A},3}}  \frac{\bfv^T\widehat{\bfSigma}_{\textrm{r}}\bfv}{\|\bfv\|_2^{2}},$$
where $$\mathcal{C}_{\mathcal{A},3}= \{\bfv \in \mathcal{R}^p: \|\bfv_{\mathcal{I}}\|_1 \leq 3\|\bfv_{\mathcal{A}}\|_1\}.$$


\begin{theorem}\label{esterr}
Assume  the rows of $\bfX$ and $\bfX_{\mathrm{r}}$ are i.i.d sub-Gaussian samples drawn  from  population with mean $\textbf{0}$ and covariance matrix $\bfSigma$, and  $\widehat{\bfSigma}_{\mathrm{r}}$ satisfying restricted eigenvalue  condition with   $\phi_{\widehat{\bfSigma}_{\mathrm{r}}} \geq \phi_0 > 0,$ and the noise vector $\bfepsilon$ is mean zero sub-Gaussian with noise level $\sigma_{\epsilon}$, and $n\geq \frac{4}{C}\log p,$ $n_{\mathrm{r}}\geq \frac{4}{C}\log p.$
Take $\lambda\geq 2 \lambda_0  = 4(\frac{C_1(C_3+C_4)}{\sqrt{C}}\sqrt{\frac{\log p}{n}}  +  \frac{C_1+\sqrt{C}C_2}{\sqrt{C}}\sigma_{\epsilon}\sqrt{\frac{\log p}{ n}} + \frac{C_1C_3}{\sqrt{C}} \sqrt{\frac{\log p}{n_{\mathrm{r}}}})$.

(i) With probability at least $1-3/p^2- 1/p^3$
we have $$\| \widehat{\bfbeta}^{\mathrm{c}}  -\bfbeta^*_{\mathcal{A}}\|  \leq \frac{6}{\phi_0}(\frac{C_1(C_3+C_4)}{\sqrt{C}}\sqrt{\frac{s\log p}{n}}  +  \frac{C_1+\sqrt{C}C_2}{\sqrt{C}}\sigma_{\epsilon}\sqrt{\frac{s\log p}{ n}} + \frac{C_1C_3}{\sqrt{C}} \sqrt{\frac{s\log p}{n_{\mathrm{r}}}}). $$

(ii)
Suppose we observe $\bfX_{\mathrm{new}} \in \mathcal{R}^{n_{\mathrm{new}}\times p}$, whose rows are sampled from the same distribution as that of $\bfX$'s. Then with probability at least $1-3/p^2- 1/p^3$,  the prediction error satisfies
$$\|\bfX_{\mathrm{new}} (\widehat{\bfbeta}^{\mathrm{c}} - \bfbeta^*_{\mathcal{A}})\|_2^2/n_{\mathrm{new}}\leq \mathcal{O}((\sigma_{\epsilon}\sqrt{\frac{s\log{p}}{n}} + \sqrt{\frac{s\log{p}}{n_{\mathrm{r}}}})^2) (1+ s^2 (\sqrt{\frac{\log p}{n_{\mathrm{r}}}}+ \sqrt{\frac{\log p}{n_{\mathrm{new}}}}) ).$$
\end{theorem}

\begin{rem}
The assumption that  $\bfX_{\mathrm{r}}$ are i.i.d sub-Gaussian samples drawn  from  population with mean $\textbf{0}$ and covariance matrix $\bfSigma$ implies the restricted eigenvalue condition  $\phi_{\widehat{\bfSigma}_{\mathrm{r}}} \geq \phi_0 $ holds  for some positive  $\phi_0$ with high probability as long as
$n_{\mathrm{r}} \geq \mathcal{O}(s\log{p})$ \cite{VanPeter:2009,Vershynin:2010,HuangJiaoLuZhu:2017}.
As shown in Theorem \ref{esterr}, with the help of reference panel,  we can also get an accurate estimator by \eqref{remi_1} even if we only have marginal information in high-dimension setting  as long as $n \geq \mathcal{O}(s\log{p})$ and $n_{\mathrm{r}} \geq \mathcal{O}(s\log{p})$.  Furthermore, the estimation error of REMI model \eqref{remi_1}
achieve the minimax optimal rate  as that of the Lasso  \cite{raskutti2011minimax} if the number of samples of reference penal $n_{\mathrm{r}}$ is at the same order of the
number of individual-level  samples $n$.
Moreover if
the magnitude of the  significant entries larger than $\mathcal{O}(\sigma_{\epsilon} \sqrt{\frac{s\log{p}}{n}} + \sqrt{\frac{s\log{p}}{n_{\mathrm{r}}}} )$, the estimated support $\mathrm{supp}(\widehat{\bfbeta}^{\mathrm{c}})$ coincide with the true significant set  $\mathcal{A}$.
\end{rem}

\section{Numerical Studies}
\label{emp_results}
\subsection{Simulation studies}
\label{simulation_study}
In simulation studies, we compare 
REMI-C~(\ref{remi_1}), REMI-R~(\ref{remi_2}) and  Lasso using individual-level data.
To avoid unrealistic LD pattern in simulation, we used the genotype data $\bfX$ from the GERA data set~\cite{hoffmann2011next}. The GERA data set provided 657,184 genotyped SNPs for 62,313 European individuals. We performed strict quality control on data using PLINK~\cite{purcell2007plink}. We excluded SNPs with a minor allele frequency less than 1$\%$, having missing values in more than 1$\%$ of the individuals or with a Hardy-Weinberg equilibrium $p$-value below 0.0001. Moreover, we removed one member of pairs with genetic relatedness larger than 0.05. Finally, there remained 53,940 samples for 550,482 SNPs.

As individual-level-based analyses often suffer from limited sample sizes due to the restricted access of individual-level data, summary-level-based analyses may have advantages because the sample sizes are often much larger. To simulate this situation, we prefixed the sample size for individual-level-based analyses at $n_{\textrm{ind}} =3,000$. Specifically we randomly selected $n_{\textrm{ind}}$ samples from 53,940 individuals in the GERA data set to form the genotype matrix $\bfX \in \mathbb{R}^{n_{\textrm{ind}}\times p}$, where $p=19,865$ was the total number of the genotyped SNP on chromosomes 16, 17 and 18. Then the phenotype vector $\bfy$ was generated as $\bfy=\bfX\bfbeta^* + \bfepsilon$, where $\bfepsilon\sim \mathcal{N}(0,\sigma^2_{\epsilon})$ and the heritability ($h^2 = \frac{\mathrm{Var}(\bfX\bfbeta) }{\mathrm{Var}(\bfX\bfbeta^*) + \sigma^2_{\epsilon}}$) was controlled at 0.2, 0.3, 0.4 and 0.5. Here $\bfbeta^*$ was the vector of true effect size with sparsity $\alpha$, i.e., $\alpha\times p$ entries in $\bfbeta^*$ were nonzero and they were sampled from $\mathcal{N}(0,1)$. In our simulation study, we varied $\alpha$ in $\{0.001, 0.003, 0.005, 0.007, 0.01, 0.02\}$. With $\{\bfX,\bfy\}$ at hand, the standard Lasso can be applied, serving as a reference of individual-level data analysis.

To generate summary-level data, we varied sample size $n$ from 3,000 to 50,000. We generated individual-level data as  we described above and then we ran simple linear regression on $\{\bfX_j,\bfy\}$, $j = 1,\dots,p$ to obtain $\{\widehat{\bfbeta}^{(m)},\hat{\bfs}^2\}$. After that, we pretended that we did not have individual-level data $\{\bfX,\bfy\}$ and then only used $\bfX^T\bfy$ and $\{\widehat{\bfbeta}^{(m)},\hat{\bfs}^2\}$ as the input for REMI-C and REMI-R, respectively. We used 379 European samples from the 1000 Genome Project data as the reference panel to estimate covariance matrix (REMI-C) or correlation matrix (REMI-R), as detailed in Section \ref{refpanel}. For each replication, we held out 200 independent samples to evaluate prediction accuracy. In total, we summarized our results based on 50 replications for each setting.

\begin{figure}[ht]
	\centering
	\includegraphics[width=1\textwidth]{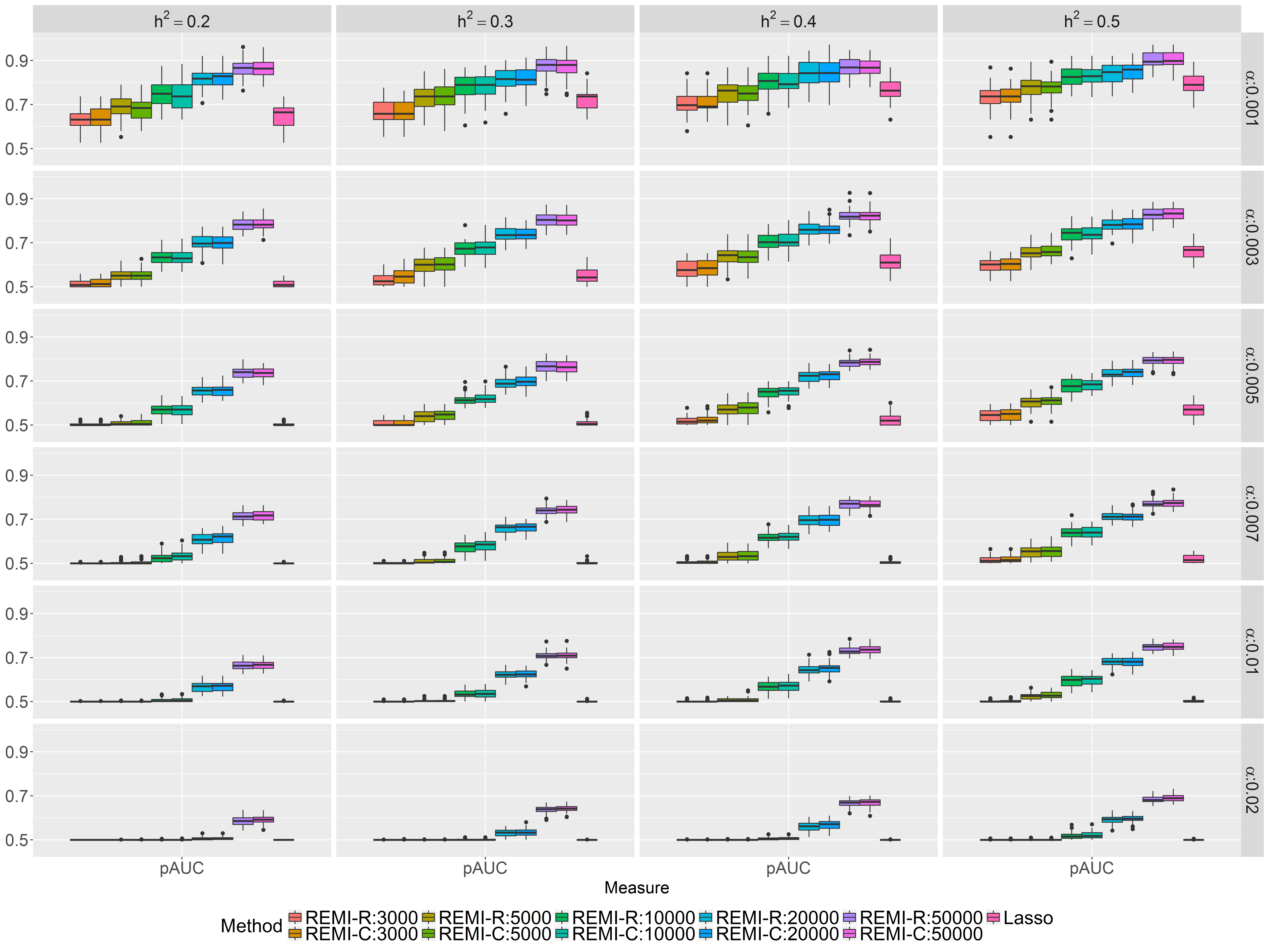}
	\caption{The comparison of variable selection performance of REMIs (REMI-R and REMI-C) for summary-statistics data with Lasso for individual-level data with sample size 3000. The sample size used to produce summary statistics was varied and denoted as $n\in\{3,000, 5,000, 10,000, 20,000, 50,000\}$. We used partial AUC to measure the variable selection performance. }
	\label{fig_selection}
\end{figure}

\begin{figure}[ht]
	\centering
	\includegraphics[width=1\textwidth]{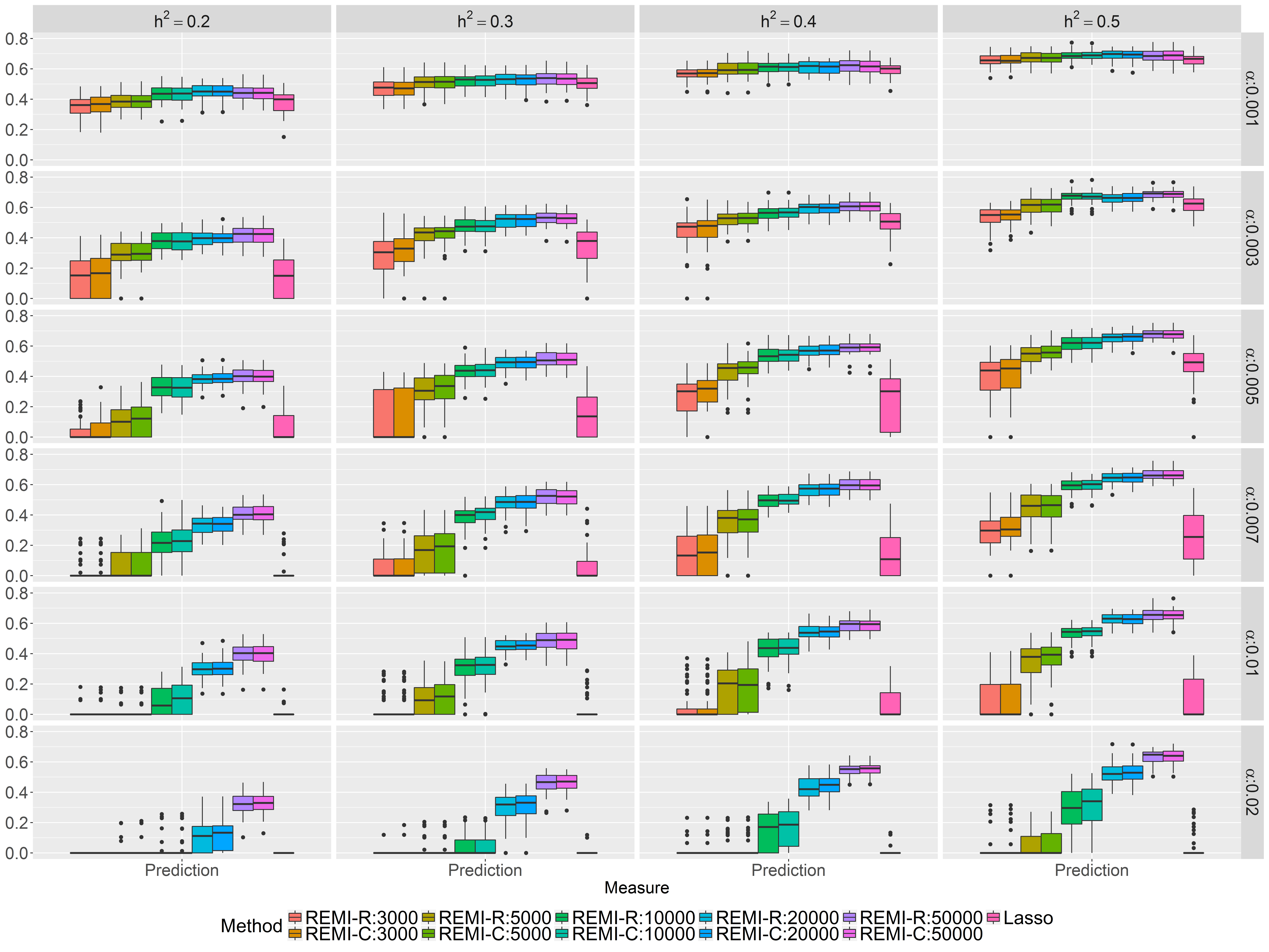}
	\caption{Prediction accuracy of REMIs with summary-level data and the Lasso for individual-level data of sample size 3,000. The sample size used to simulate summary statistics was varied $n\in\{3,000, 5,000, 10,000, 20,000, 50,000\}$. Prediction accuracy is measured by the Pearson's correlation coefficient between predicted and observed phenotypes.}
	\label{fig_pred}
\end{figure}

We compared the performance of   REMI-C, REMI-R and the Lasso using individual-level data in terms of variable selection and prediction. Specifically, we used partial area under the receiver operating characteristic (ROC) curve (partial AUC) for variable selection performance and the Pearson's correlation coefficient between predicted and observed phenotypes for prediction performance.
The results of this simulation study are shown in Figure~\ref{fig_selection} and Figure~\ref{fig_pred}. First, we can observe that the difference between REMI-C and REMI-R is nearly invisible. This justifies the approximation made in REMI-R. Second, when the sample size ($n$ = 3,000 or 5,000) in summary-level data is similar to that of individual-level data ($n_{\mathrm{ind}} = 3,000$), the performance of variable selection and prediction for REMIs (both REMI-C and REMI-R) is similar to that of Lasso. 
Third, REMIs gradually outperform the Lasso as the sample size increases from 5,000 to 50,000, for both variable selection and prediction. This clearly indicates that REMIs can have big advantages over the Lasso when the sample sizes of summary-level data become much larger.


\subsection{Real data analysis}
\label{realdata}

\begin{figure}[ht]
	\centering
	\includegraphics[width=0.48\textwidth]{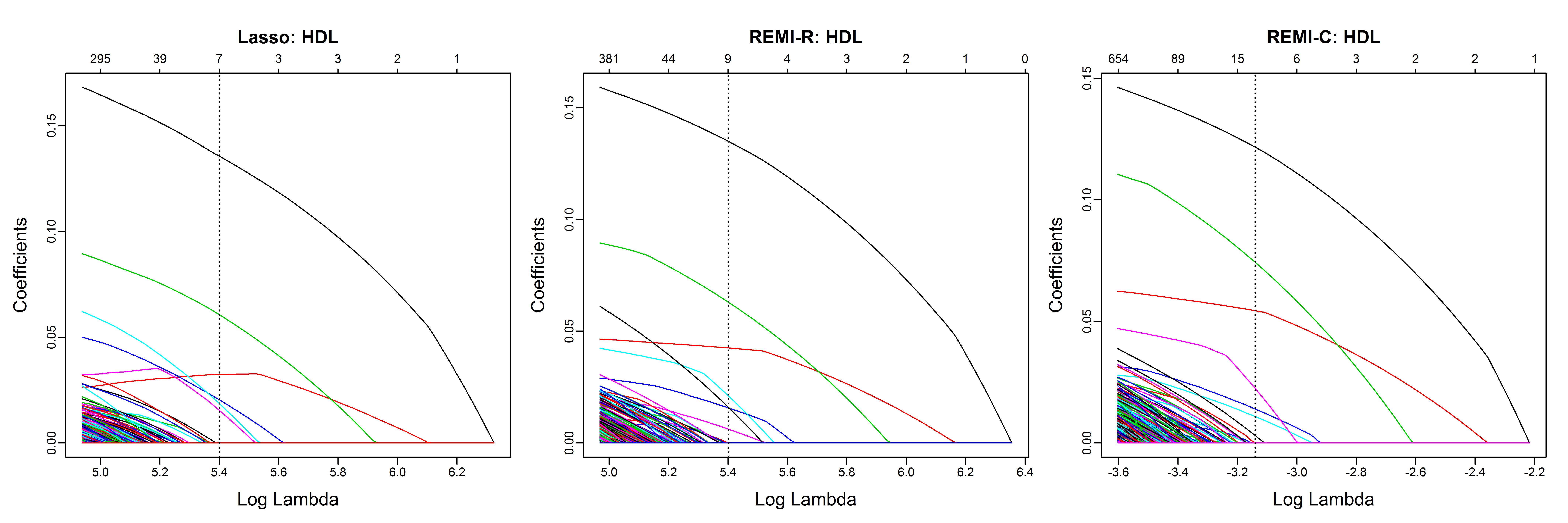}
	\includegraphics[width=0.48\textwidth]{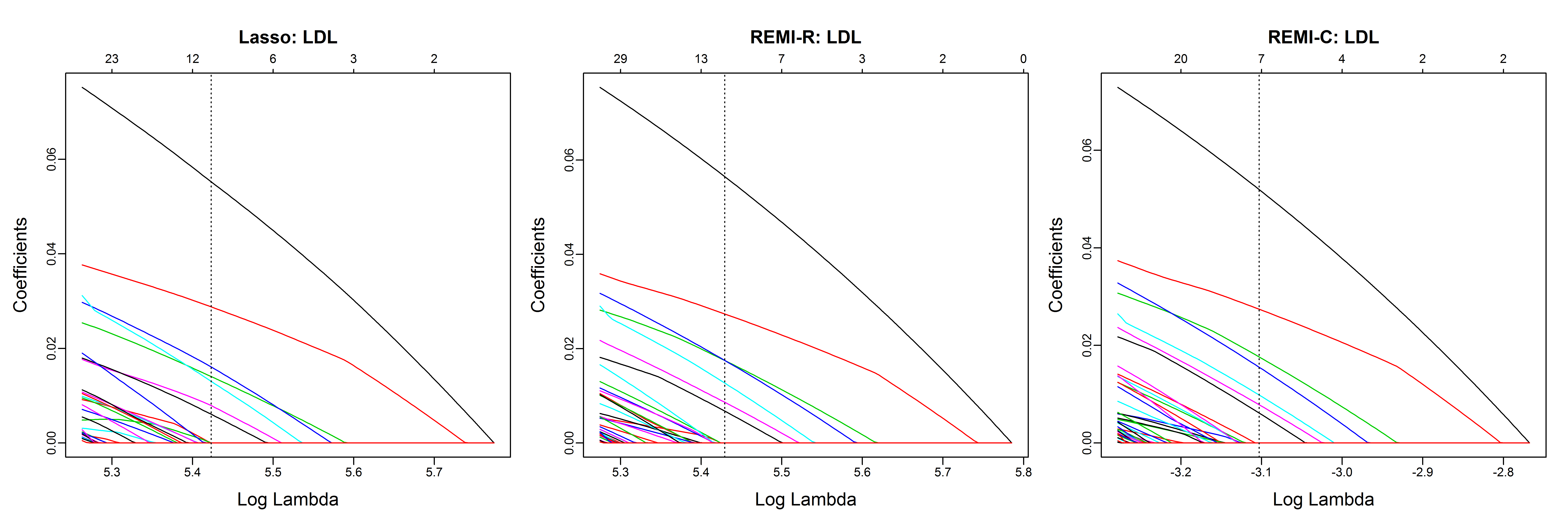}
	\includegraphics[width=0.48\textwidth]{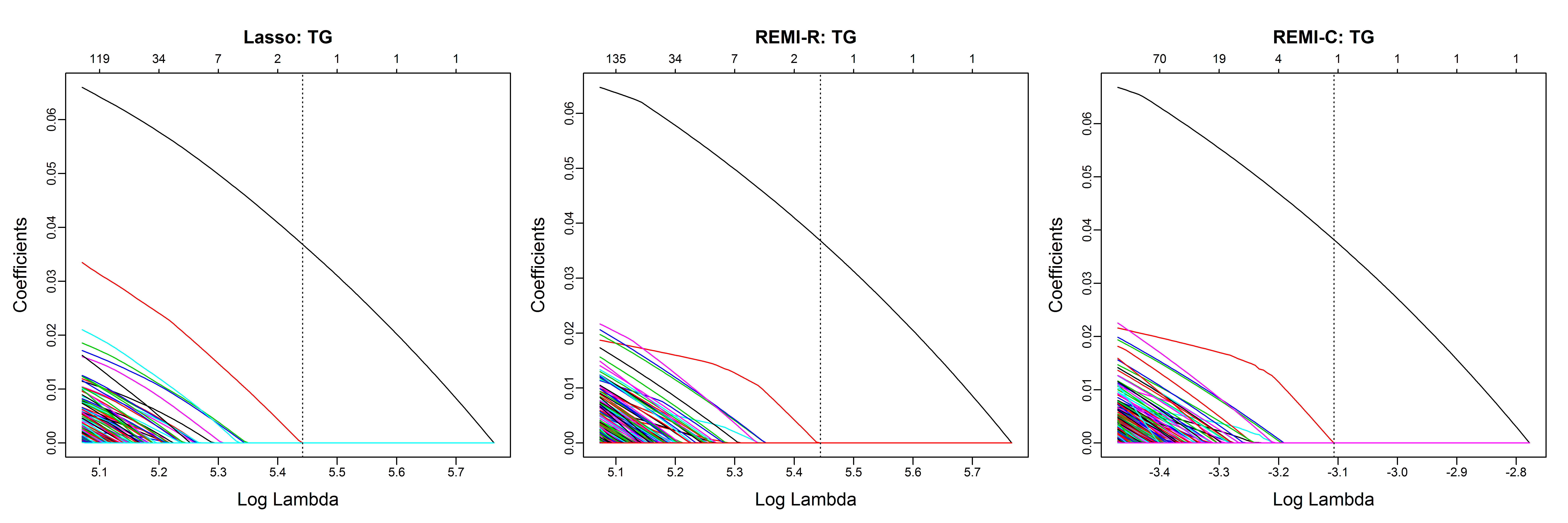}
	\includegraphics[width=0.48\textwidth]{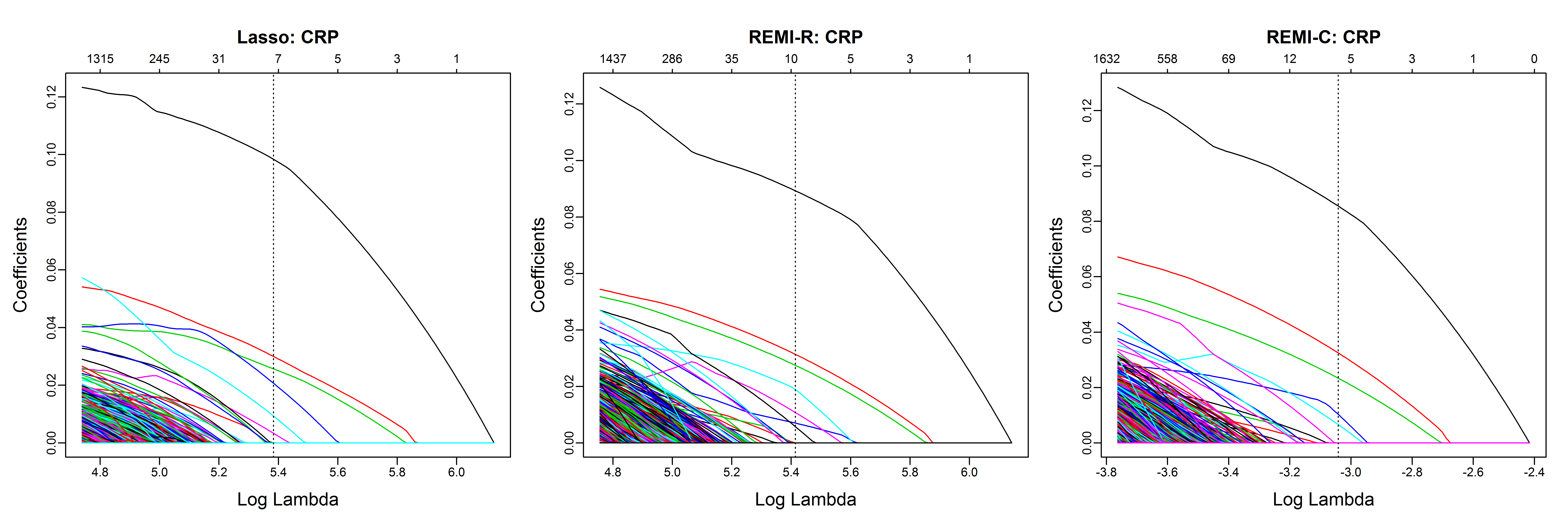}
	\includegraphics[width=0.48\textwidth]{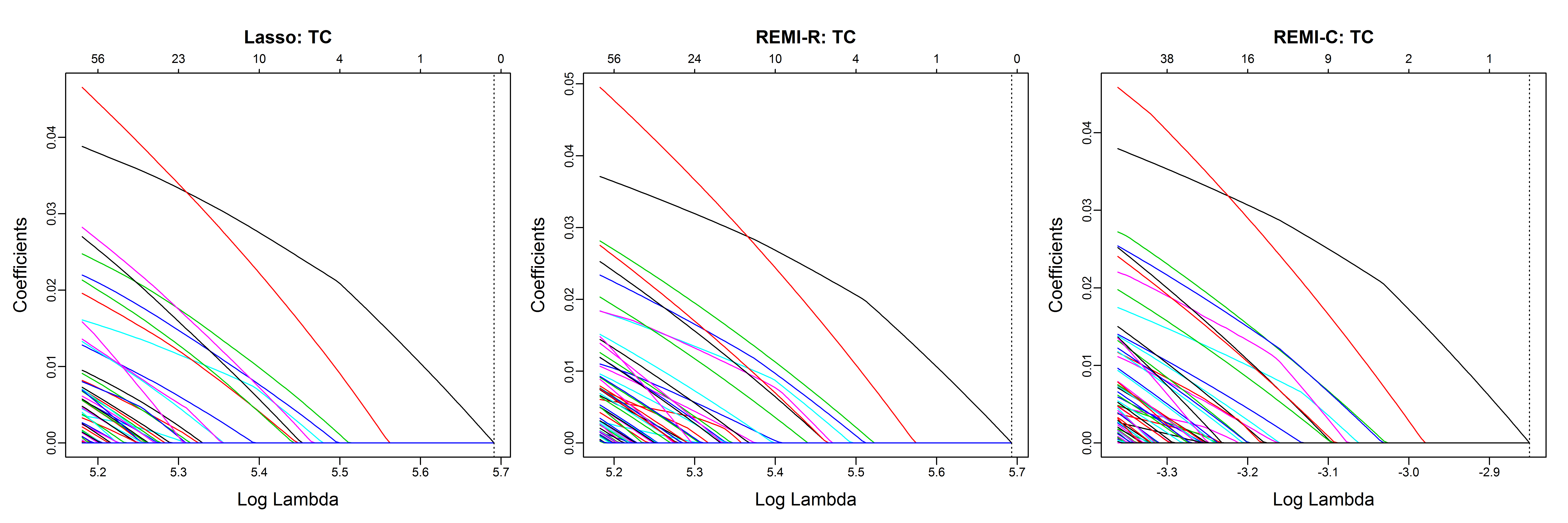}
	\includegraphics[width=0.48\textwidth]{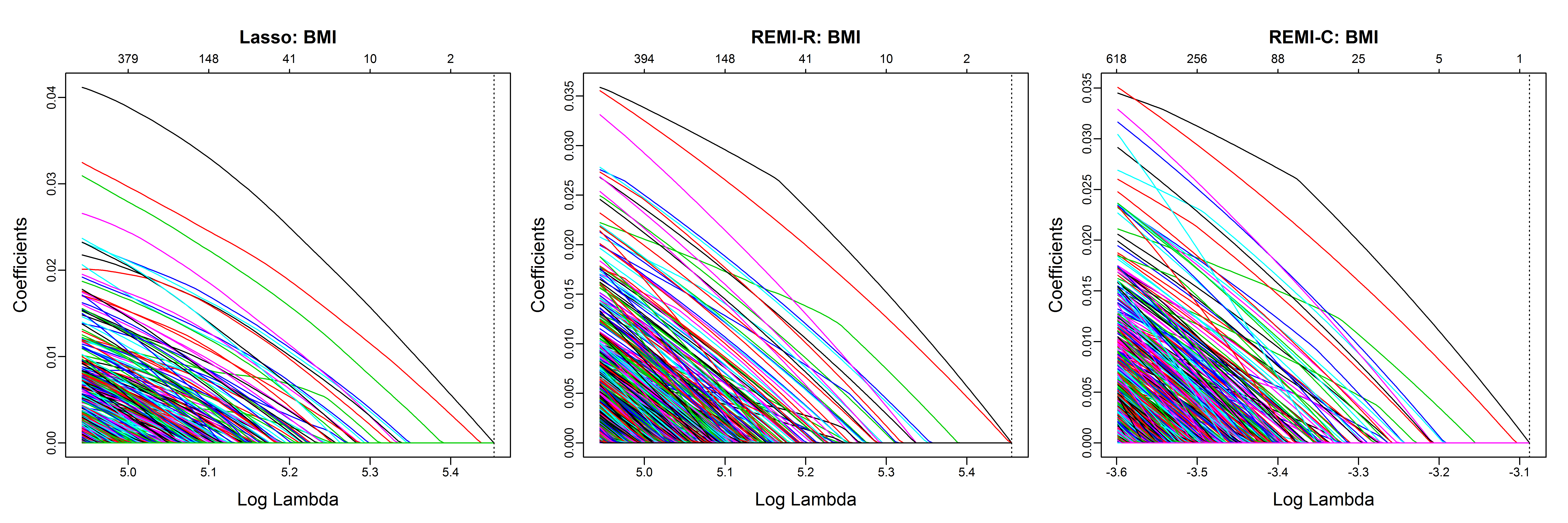}
	\includegraphics[width=0.48\textwidth]{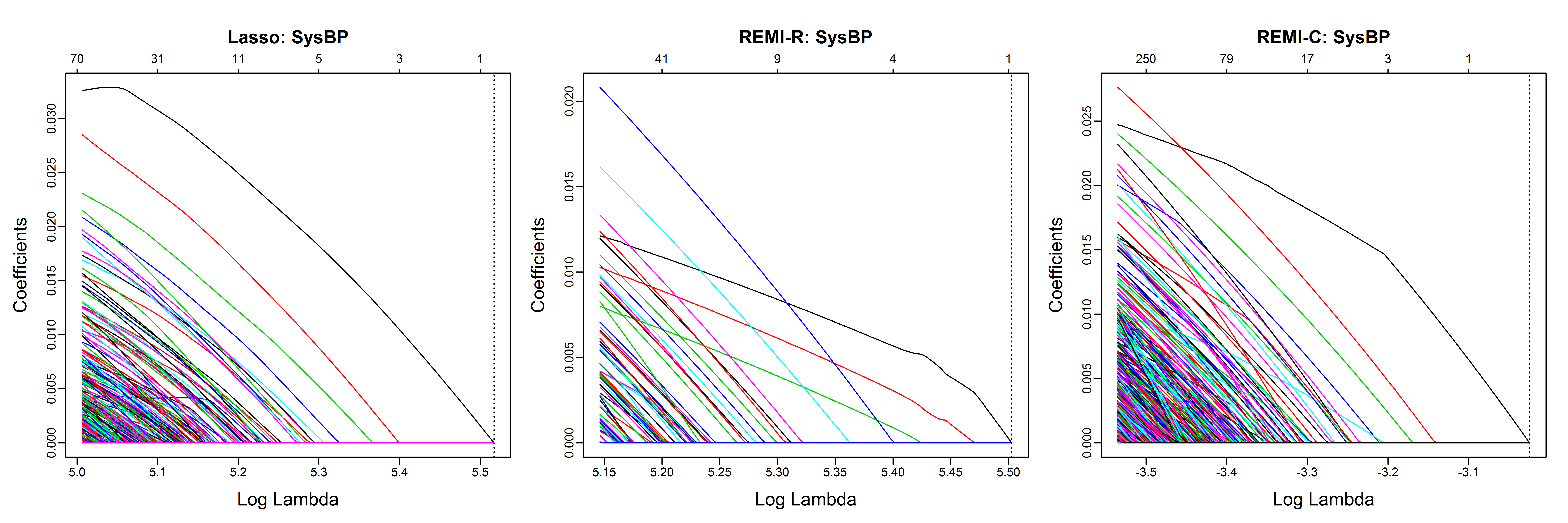}
	\includegraphics[width=0.48\textwidth]{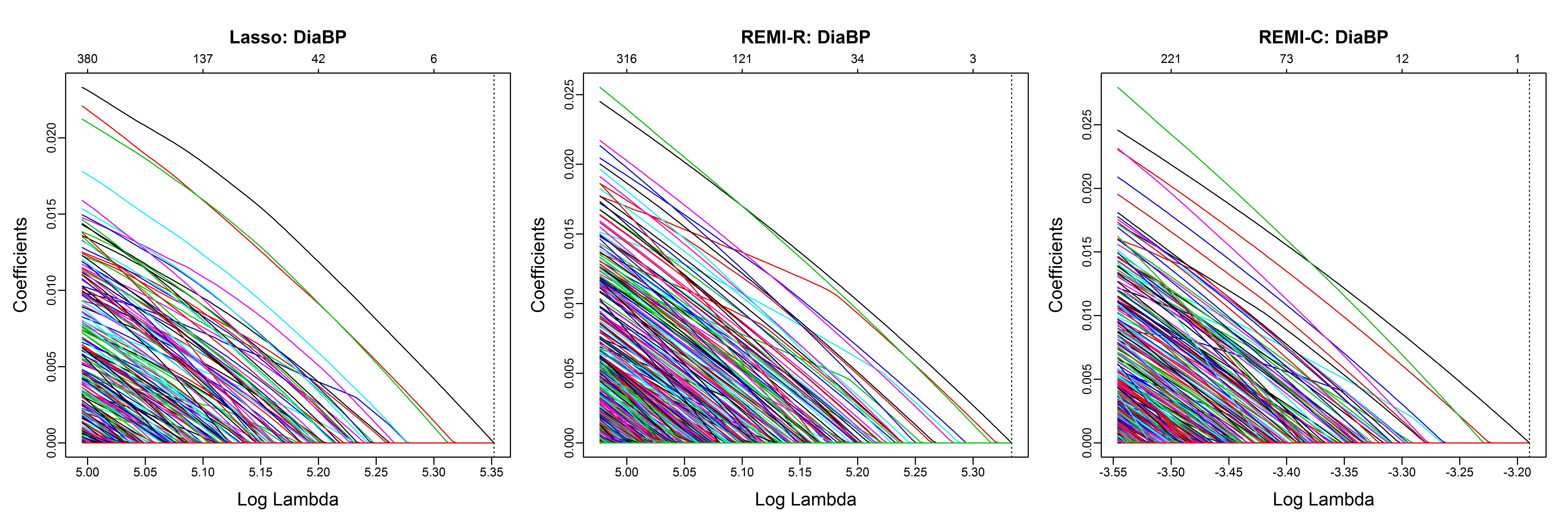}
	\includegraphics[width=0.48\textwidth]{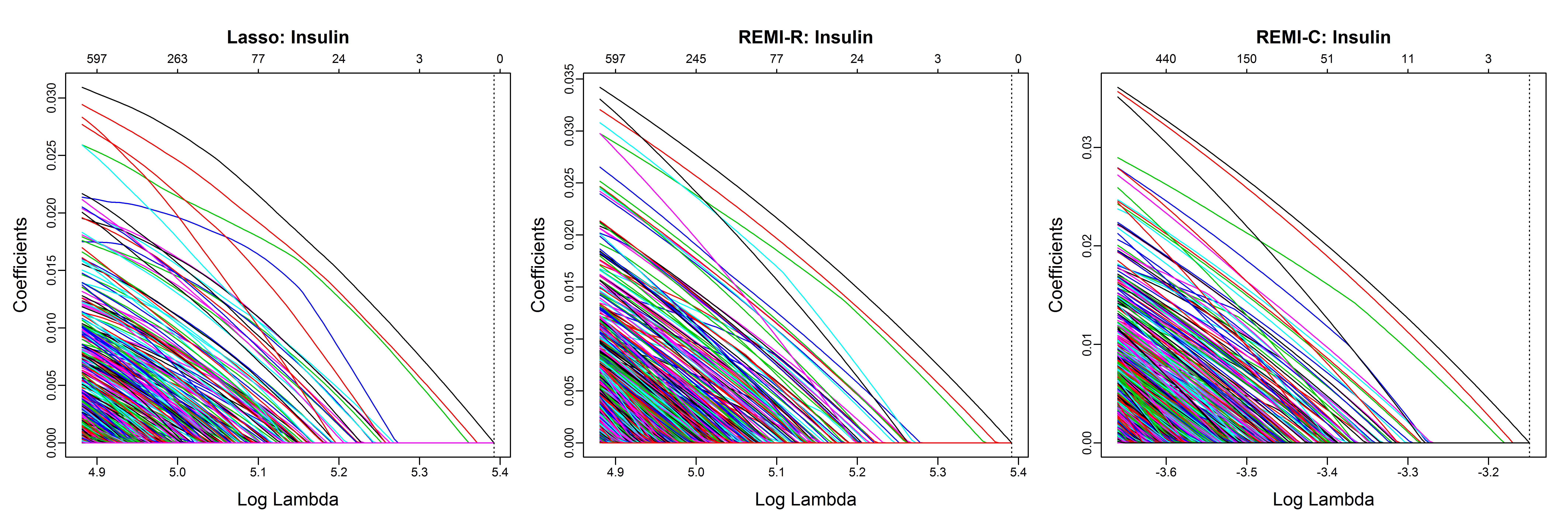}
	\includegraphics[width=0.48\textwidth]{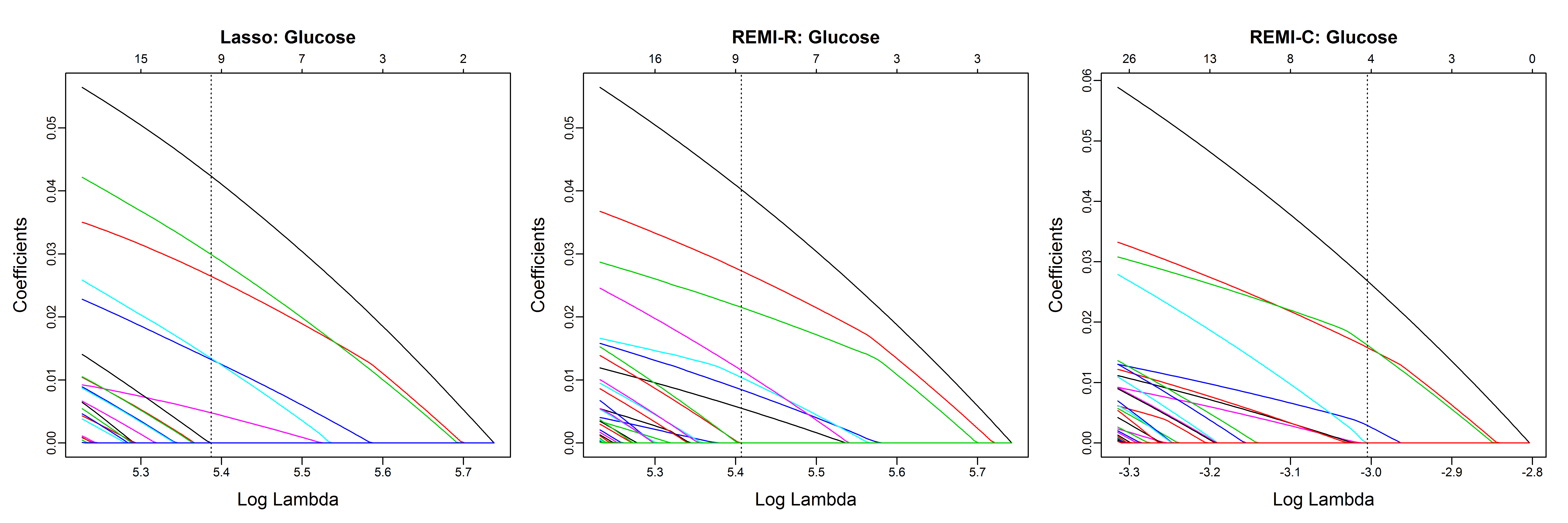}

	\caption{Solution paths of Lasso, REMI-R, and REMI-C for HDL, LDL, TG, CRP TC, BMI, SysBP, DiaBP, Insulin and Glucose using the NFBC1966 data sets.}
	\label{fig3}
\end{figure}

To demonstrate the utility of REMIs, we first compared Lasso and REMIs based on the GWAS data set from the Northern Finland Birth Cohorts program (NFBC1966)~\cite{sabatti2009genome}. The NFBC1966 data set contains information for 5,402 individuals with a selected list of phenotypic data related to cardiovascular disease including high-density lipoprotein (HDL), low-density lipoprotein (LDL), total cholesterol (TC), triglycerides (TG),  C-reactive protein (CRP), glucose, insulin, body mass index (BMI), systolic (SysBP) and diastolic (DiaBP) blood pressure. For each individual, 364,590 SNPs have been genotyped. We performed strict quality control on data using PLINK~\cite{purcell2007plink}. We first excluded individuals having discrepancies between reported sex and sex determined from the X chromosome. We also excluded SNPs with a minor allele frequency less than 1$\%$, having missing values in more than 1$\%$ of the individuals or with a Hardy-Weinberg equilibrium $p$-value below 0.0001. In particular, we selected well-imputed variants from HapMap 3 reference panel~\cite{international2010integrating}. After the strict quality control, 5,123 individuals with 310,975 SNPs in NFBC1966 were remaining for the further analysis. As we have the individual-level data, it is possible to run REMI-C, REMI-R and Lasso for all these traits. The solution paths using Lasso, REMI-R, and REMI-C for the ten metabolic traits in NFBC1966 data set are presented in Figures~\ref{fig3}. The dotted vertical bars in these two figures indicate the corresponding selected tuning parameters based on BIC. One can see that the differences among solution paths for the Lasso using individual-level data,  REMI-C and REMI-R using summary statistics are very minor, which is consistent with results from our simulation studies in Section~\ref{simulation_study}.

\begin{figure}[ht]
	\centering
	\includegraphics[width=1\textwidth]{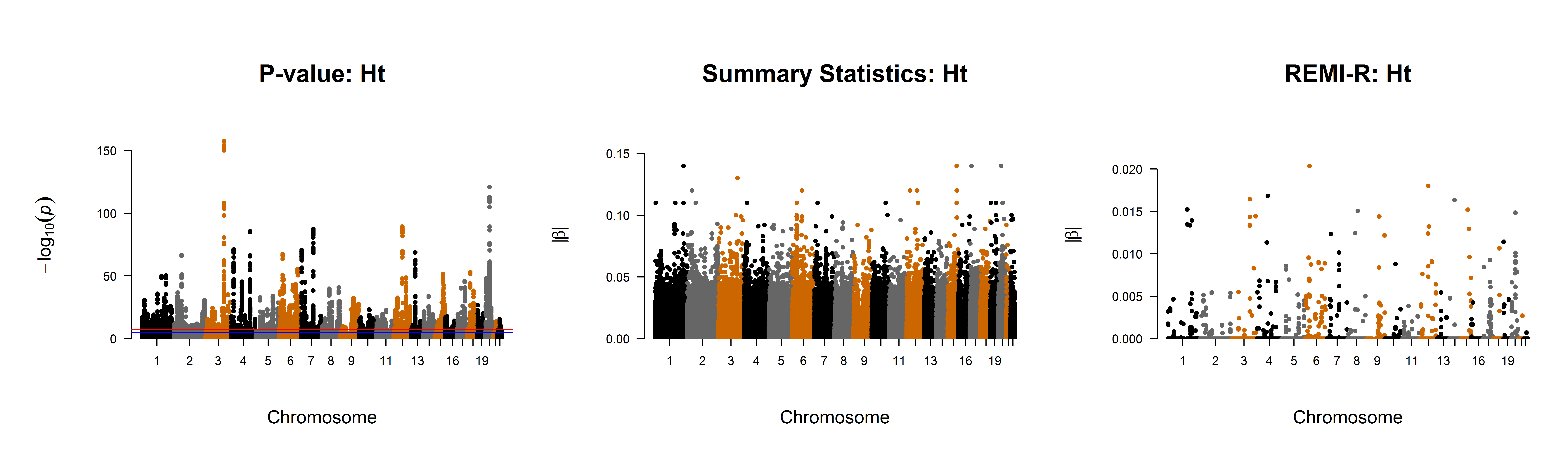}
	\caption{Manhattan plots of  analysis result of human height for: -$\mathrm{log}_{10} p$-value, $|\widehat\bfbeta^{\textrm{m}}|$ from marginal analysis and $|\widehat\bfbeta^{\textrm{r}}|$ from REMI-R.}
	\label{fig1}
\end{figure}

\begin{figure}[ht]
	\centering
	\includegraphics[width=1\textwidth]{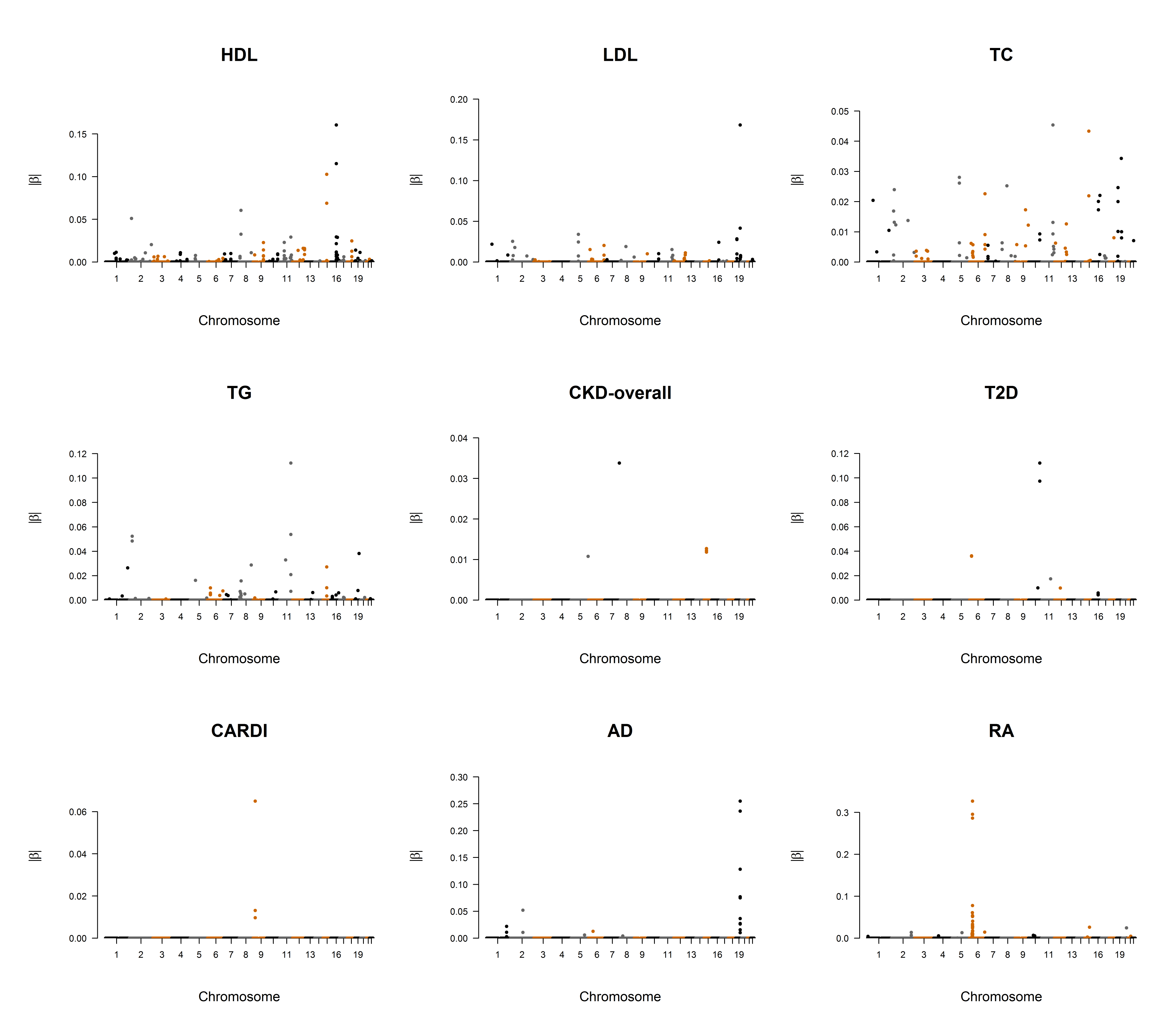}
	
	\caption{Manhattan plots of $|\widehat\bfbeta^{\textrm{r}}|$ from REMI-R for HDL, LDL, TC, TG, CKD-overall, T2D, CARDI, AD and RA.}
	\label{fig2}
\end{figure}

In the released GWAS summary-level data sets, it is often the case that $\{\widehat{\bfbeta}^{(m)},\hat{\bfs}^2\}$ rather than the inner product $\bfX^T\bfy$ is made available. Therefore, we applied REMI-R to analyze summary statistics for ten GWASs of complex phenotypes. The source of the GWASs is given in Table~\ref{Tab:01}. Because the individuals of summary-level data sets were all from European ancestry, we used 379 European-ancestry samples in 1000 Genome Project~\cite{10002012integrated} as a reference panel to estimate correlation matrix. Due to the quality of SNPs in the summary statistics, we restricted our analysis to a set of common and well-imputed variants from the HapMap 3 reference panel~\cite{international2010integrating}, which included 1,197,724 SNPs in total.
Figure~\ref{fig1} shows the Manhattan plots of summary statistics for height (Ht) including $-\log_{10}$($p$-value), $|\widehat\bfbeta^{\textrm{m}}|$ and $|\widehat\bfbeta^{\textrm{r}}|$.  The Manhattan plots of the absolute effect sizes from REMI-R for all other nine traits are shown in Figure~\ref{fig2}.

\begin{figure}[ht]
\centering 
\includegraphics[width=1\textwidth]{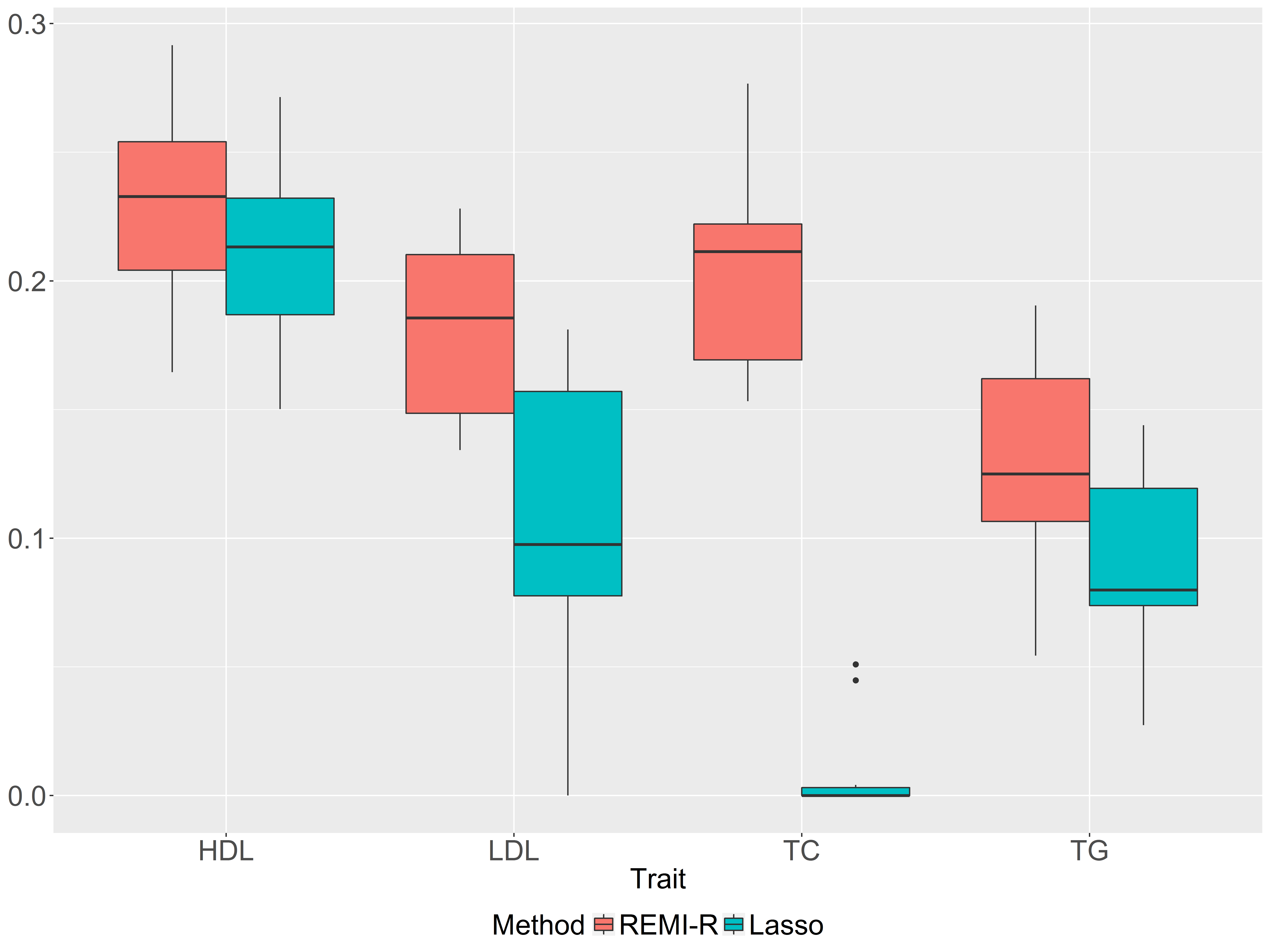}
\caption{Prediction accuracy (measured by the Pearson's correlation coefficients) of REMI-R and the Lasso for HDL, LDL, TC and TG in the NFBC1966 data sets, where REMI-R was fitted using independent summary-level data and the Lasso was fitted using the individual-level data from NFBC1966. Their prediction accuracies were evaluated on 1/10 of the NFBC1966 data set holded out for testing.}
\label{fig5}
\end{figure}

Besides the effect size estimation, we evaluated prediction performance using 5,123 samples from the NFBC1966~\cite{sabatti2009genome}. To make a fair comparison with Lasso, we first split all 5,123 samples into ten folds. On the one hand, we applied REMI-R on the summary statistics for these lipid traits listed in Table~\ref{Tab:01}. Again, we used 379 European-ancestry samples from the 1000 Genome Project as a reference panel. For each of 10 folds in NFBC1966 data set, we calculated the predicted phenotypic values and evaluated the Pearson's correlation coefficients between the predicted phenotypic values and the observed ones.  On the other hand, we fitted the Lasso on the individual-level NFBC1966 data using the same ten-fold data for cross-validation. Specifically, we randomly selected nine folds of individual-level data as the training set to fit the Lasso, and evaluated prediction accuracy of the fitted model using the remaining one fold. Note that we used the same remaining fold to evaluate the prediction accuracy of the fitted REMI-R model. The prediction performance for REMI-R and Lasso is shown in Figure~\ref{fig5}. Clearly, the prediction performance of REMI-R outperforms the standard Lasso as the sample size in the summary statistics for these lipid traits are around 100,000 but the individual-level data contains only 5,123$\times 9/10$ samples. These real data results indicate the great advantage of REMI over the Lasso for risk prediction.

%

\begin{thebibliography}{10}

\bibitem{berisa2016approximately}
T.~Berisa and J.~K. Pickrell.
\newblock Approximately independent linkage disequilibrium blocks in human
  populations.
\newblock {\em Bioinformatics (Oxford, England)}, 32(2):283, 2016.

\bibitem{BickelRitovTsybakov:2009}
P.~J. Bickel, Y.~Ritov, and A.~B. Tsybakov.
\newblock {Simultaneous analysis of Lasso and Dantzig selector}.
\newblock {\em Ann. Statist.}, 37(4):1705--1732, 2009.

\bibitem{Bickel:2009}
P.~J. Bickel, Y.~Ritov, and A.~B. Tsybakov.
\newblock Simultaneous analysis of lasso and dantzig selector.
\newblock {\em The Annals of Statistics}, pages 1705--1732, 2009.

\bibitem{buhlmann2011statistics}
P.~B{\"u}hlmann and S.~Van De~Geer.
\newblock {\em Statistics for high-dimensional data: methods, theory and
  applications}.
\newblock Springer, 2011.

\bibitem{bulik2015ld}
B.~K. Bulik-Sullivan, P.-R. Loh, H.~K. Finucane, S.~Ripke, J.~Yang,
  N.~Patterson, M.~J. Daly, A.~L. Price, B.~M. Neale, S.~W.~G. of~the
  Psychiatric Genomics~Consortium, et~al.
\newblock Ld score regression distinguishes confounding from polygenicity in
  genome-wide association studies.
\newblock {\em Nature genetics}, 47(3):291--295, 2015.

\bibitem{CandesTao:2007}
E.~Candes, T.~Tao, et~al.
\newblock The dantzig selector: Statistical estimation when p is much larger
  than n.
\newblock {\em The Annals of Statistics}, 35(6):2313--2351, 2007.

\bibitem{chen2015fine}
W.~Chen, B.~R. Larrabee, I.~G. Ovsyannikova, R.~B. Kennedy, I.~H. Haralambieva,
  G.~A. Poland, and D.~J. Schaid.
\newblock Fine mapping causal variants with an approximate bayesian method
  using marginal test statistics.
\newblock {\em Genetics}, 200(3):719--736, 2015.

\bibitem{10002012integrated}
.~G.~P. Consortium et~al.
\newblock An integrated map of genetic variation from 1,092 human genomes.
\newblock {\em Nature}, 491(7422):56, 2012.

\bibitem{international2010integrating}
I.~H.~. Consortium et~al.
\newblock Integrating common and rare genetic variation in diverse human
  populations.
\newblock {\em Nature}, 467(7311):52, 2010.

\bibitem{FanLi:2001}
J.~Fan and R.~Li.
\newblock Variable selection via nonconcave penalized likelihood and its oracle
  properties.
\newblock {\em J. Amer. Statist. Assoc.}, 96(456):1348--1360, 2001.

\bibitem{friedman2010regularization}
J.~Friedman, T.~Hastie, and R.~Tibshirani.
\newblock Regularization paths for generalized linear models via coordinate
  descent.
\newblock {\em Journal of statistical software}, 33(1):1, 2010.

\bibitem{fuchsberger2016genetic}
C.~Fuchsberger, J.~Flannick, T.~M. Teslovich, A.~Mahajan, V.~Agarwala, K.~J.
  Gaulton, C.~Ma, P.~Fontanillas, L.~Moutsianas, D.~J. McCarthy, et~al.
\newblock The genetic architecture of type 2 diabetes.
\newblock {\em Nature}, 2016.

\bibitem{gratten2014large}
J.~Gratten, N.~R. Wray, M.~C. Keller, and P.~M. Visscher.
\newblock Large-scale genomics unveils the genetic architecture of psychiatric
  disorders.
\newblock {\em Nature neuroscience}, 17(6):782--790, 2014.

\bibitem{Hastie2009elements}
T.~Hastie, R.~Tibshirani, and J.~Friedman.
\newblock {\em The elements of statistical learning (2nd Edition)}.
\newblock Springer, 2009.

\bibitem{hastie2015statistical}
T.~Hastie, R.~Tibshirani, and M.~Wainwright.
\newblock {\em Statistical learning with sparsity: the lasso and
  generalizations}.
\newblock CRC press, 2015.

\bibitem{hindorff2009potential}
L.~A. Hindorff, P.~Sethupathy, H.~A. Junkins, E.~M. Ramos, J.~P. Mehta, F.~S.
  Collins, and T.~A. Manolio.
\newblock Potential etiologic and functional implications of genome-wide
  association loci for human diseases and traits.
\newblock {\em Proceedings of the National Academy of Sciences},
  106(23):9362--9367, 2009.

\bibitem{hoffmann2011next}
T.~J. Hoffmann, M.~N. Kvale, S.~E. Hesselson, Y.~Zhan, C.~Aquino, Y.~Cao,
  S.~Cawley, E.~Chung, S.~Connell, J.~Eshragh, et~al.
\newblock Next generation genome-wide association tool: design and coverage of
  a high-throughput european-optimized snp array.
\newblock {\em Genomics}, 98(2):79--89, 2011.

\bibitem{hormozdiari2014identifying}
F.~Hormozdiari, E.~Kostem, E.~Y. Kang, B.~Pasaniuc, and E.~Eskin.
\newblock Identifying causal variants at loci with multiple signals of
  association.
\newblock {\em Genetics}, 198(2):497--508, 2014.

\bibitem{hu2017leveraging}
Y.~Hu, Q.~Lu, R.~Powles, X.~Yao, C.~Yang, F.~Fang, X.~Xu, and H.~Zhao.
\newblock Leveraging functional annotations in genetic risk prediction for
  human complex diseases.
\newblock {\em PLOS Computational Biology}, 13(6):e1005589, 2017.

\bibitem{HuangJiaoLuZhu:2017}
J.~Huang, Y.~Jiao, X.~Lu, and L.~Zhu.
\newblock Robust decoding from 1-bit compressive sampling with least squares.
\newblock {\em arXiv preprint arXiv:1711.01206}, 2017.

\bibitem{kichaev2014integrating}
G.~Kichaev, W.-Y. Yang, S.~Lindstrom, F.~Hormozdiari, E.~Eskin, A.~L. Price,
  P.~Kraft, and B.~Pasaniuc.
\newblock Integrating functional data to prioritize causal variants in
  statistical fine-mapping studies.
\newblock {\em PLoS Genet}, 10(10):e1004722, 2014.

\bibitem{lee2011estimating}
S.~H. Lee, N.~R. Wray, M.~E. Goddard, and P.~M. Visscher.
\newblock Estimating missing heritability for disease from genome-wide
  association studies.
\newblock {\em The American Journal of Human Genetics}, 88(3):294--305, 2011.

\bibitem{MeinshausenBuhlmann:2006}
N.~Meinshausen and P.~B{\"u}hlmann.
\newblock High-dimensional graphs and variable selection with the lasso.
\newblock {\em Ann. Statist.}, 34(3):1436--1462, 2006.

\bibitem{editor2012ng}
E.~of~Nature~Genetics.
\newblock Asking for more.
\newblock {\em Nature Genetics}, 44:733, 2012.

\bibitem{pasaniuc2014fast}
B.~Pasaniuc, N.~Zaitlen, H.~Shi, G.~Bhatia, A.~Gusev, J.~Pickrell,
  J.~Hirschhorn, D.~P. Strachan, N.~Patterson, and A.~L. Price.
\newblock Fast and accurate imputation of summary statistics enhances evidence
  of functional enrichment.
\newblock {\em Bioinformatics}, 30(20):2906--2914, 2014.

\bibitem{pickrell2014joint}
J.~K. Pickrell.
\newblock Joint analysis of functional genomic data and genome-wide association
  studies of 18 human traits.
\newblock {\em The American Journal of Human Genetics}, 94(4):559--573, 2014.

\bibitem{purcell2007plink}
S.~Purcell, B.~Neale, K.~Todd-Brown, L.~Thomas, M.~A. Ferreira, D.~Bender,
  J.~Maller, P.~Sklar, P.~I. De~Bakker, M.~J. Daly, et~al.
\newblock {PLINK}: a tool set for whole-genome association and population-based
  linkage analyses.
\newblock {\em The American Journal of Human Genetics}, 81(3):559--575, 2007.

\bibitem{raskutti2011minimax}
G.~Raskutti, M.~J. Wainwright, and B.~Yu.
\newblock Minimax rates of estimation for high-dimensional linear regression
  over $\ell\_q $-balls.
\newblock {\em IEEE transactions on information theory}, 57(10):6976--6994,
  2011.

\bibitem{sabatti2009genome}
C.~Sabatti, A.-L. Hartikainen, A.~Pouta, S.~Ripatti, J.~Brodsky, C.~G. Jones,
  N.~A. Zaitlen, T.~Varilo, M.~Kaakinen, U.~Sovio, et~al.
\newblock Genome-wide association analysis of metabolic traits in a birth
  cohort from a founder population.
\newblock {\em Nature genetics}, 41(1):35, 2009.

\bibitem{schafer2005shrinkage}
J.~Sch{\"a}fer, K.~Strimmer, et~al.
\newblock A shrinkage approach to large-scale covariance matrix estimation and
  implications for functional genomics.
\newblock {\em Statistical applications in genetics and molecular biology},
  4(1):32, 2005.

\bibitem{Tibshirani:1996}
R.~Tibshirani.
\newblock Regression shrinkage and selection via the lasso.
\newblock {\em J. Roy. Statist. Soc. Ser. B}, 58(1):267--288, 1996.

\bibitem{VanPeter:2009}
S.~A. Van De~Geer, P.~B{\"u}hlmann, et~al.
\newblock On the conditions used to prove oracle results for the lasso.
\newblock {\em Electronic Journal of Statistics}, 3:1360--1392, 2009.

\bibitem{Vershynin:2016}
R.~Vershynin.
\newblock High dimensional probability.

\bibitem{Vershynin:2010}
R.~Vershynin.
\newblock Introduction to the non-asymptotic analysis of random matrices.
\newblock {\em arXiv preprint arXiv:1011.3027}, 2010.

\bibitem{vilhjalmsson2015modeling}
B.~J. Vilhj{\'a}lmsson, J.~Yang, H.~K. Finucane, A.~Gusev, S.~Lindstr{\"o}m,
  S.~Ripke, G.~Genovese, P.-R. Loh, G.~Bhatia, R.~Do, et~al.
\newblock Modeling linkage disequilibrium increases accuracy of polygenic risk
  scores.
\newblock {\em The American Journal of Human Genetics}, 97(4):576--592, 2015.

\bibitem{visscher2012five}
P.~M. Visscher, M.~A. Brown, M.~I. McCarthy, and J.~Yang.
\newblock Five years of gwas discovery.
\newblock {\em The American Journal of Human Genetics}, 90(1):7--24, 2012.

\bibitem{visscher2008heritability}
P.~M. Visscher, W.~G. Hill, and N.~R. Wray.
\newblock Heritability in the genomics era--concepts and misconceptions.
\newblock {\em Nature Reviews Genetics}, 9(4):255--266, 2008.

\bibitem{visscher201710}
P.~M. Visscher, N.~R. Wray, Q.~Zhang, P.~Sklar, M.~I. McCarthy, M.~A. Brown,
  and J.~Yang.
\newblock {10 years of GWAS discovery: biology, function, and translation}.
\newblock {\em The American Journal of Human Genetics}, 101(1):5--22, 2017.

\bibitem{welter2014nhgri}
D.~Welter, J.~MacArthur, J.~Morales, T.~Burdett, P.~Hall, H.~Junkins, A.~Klemm,
  P.~Flicek, T.~Manolio, L.~Hindorff, et~al.
\newblock {The NHGRI GWAS Catalog, a curated resource of SNP-trait
  associations}.
\newblock {\em Nucleic Acids Research}, 42(D1):D1001--D1006, 2014.

\bibitem{wood2014defining}
A.~R. Wood, T.~Esko, J.~Yang, S.~Vedantam, T.~H. Pers, S.~Gustafsson, A.~Y.
  Chu, K.~Estrada, J.~Luan, Z.~Kutalik, et~al.
\newblock Defining the role of common variation in the genomic and biological
  architecture of adult human height.
\newblock {\em Nature Genetics}, 46(11):1173--1186, 2014.

\bibitem{yang2015genome}
J.~Yang, A.~Bakshi, Z.~Zhu, G.~Hemani, A.~A. Vinkhuyzen, I.~M. Nolte, J.~V. van
  Vliet-Ostaptchouk, H.~Snieder, T.~Esko, L.~Milani, et~al.
\newblock Genome-wide genetic homogeneity between sexes and populations for
  human height and body mass index.
\newblock {\em Human molecular genetics}, page ddv443, 2015.

\bibitem{yang2010common}
J.~Yang, B.~Benyamin, B.~P. McEvoy, S.~Gordon, A.~K. Henders, D.~R. Nyholt,
  P.~A. Madden, A.~C. Heath, N.~G. Martin, G.~W. Montgomery, et~al.
\newblock Common snps explain a large proportion of the heritability for human
  height.
\newblock {\em Nature genetics}, 42(7):565--569, 2010.

\bibitem{yang2011gcta}
J.~Yang, S.~H. Lee, M.~E. Goddard, and P.~M. Visscher.
\newblock {GCTA: a tool for genome-wide complex trait analysis}.
\newblock {\em The American Journal of Human Genetics}, 88(1):76--82, 2011.

\bibitem{Zhang:2010a}
C.-H. Zhang.
\newblock Nearly unbiased variable selection under minimax concave penalty.
\newblock {\em Ann. Statist.}, 38(2):894--942, 2010.

\bibitem{ZhangHuang:2008}
C.-H. Zhang and J.~Huang.
\newblock The sparsity and bias of the {LASSO} selection in high-dimensional
  linear regression.
\newblock {\em Ann. Statist.}, 36(4):1567--1594, 2008.

\bibitem{ZhangZhang:2012}
C.-H. Zhang and T.~Zhang.
\newblock A general theory of concave regularization for high-dimensional
  sparse estimation problems.
\newblock {\em Statist. Sci.}, 27(4):576--593, 2012.

\bibitem{ZhaoYu:2006}
P.~Zhao and B.~Yu.
\newblock On model selection consistency of {L}asso.
\newblock {\em J. Mach. Learn. Res.}, 7:2541--2563, 2006.

\bibitem{zhou2016unified}
X.~Zhou.
\newblock A unified framework for variance component estimation with summary
  statistics in genome-wide association studies.
\newblock {\em bioRxiv}, page 042846, 2016.

\bibitem{zhu2016bayesian}
X.~Zhu and M.~Stephens.
\newblock Bayesian large-scale multiple regression with summary statistics from
  genome-wide association studies.
\newblock {\em bioRxiv}, page 042457, 2016.

\bibitem{zou2007degrees}
H.~Zou, T.~Hastie, R.~Tibshirani, et~al.
\newblock On the degrees of freedom of the lasso.
\newblock {\em The Annals of Statistics}, 35(5):2173--2192, 2007.

\end{thebibliography}

\begin{landscape}
\begin{table}[ht]
\tiny
\caption{GWAS data sets in our experiment} \label{Tab:01}%
\centering
\begin{tabular}{l l l l l l}
\hline
  ID & YEAR & Traits & Sample Size & SNPs & Link  \\
  \hline
  AD & 2013 & Alzheimer Disorder & 54162& 1149751 & \url{http://www.pasteur-lille.fr/en/recherche/u744/igap/igap_download.php} \\
  CARDI & 2015 & Coronary Artery Disease &817857& 1197724 & \url{http://www.cardiogramplusc4d.org/data-downloads/}\\
  CKD-overall & 2015 & eGFRcrea in overall population & 133715& 984086 & \url{https://www.nhlbi.nih.gov/research/intramural/researchers/ckdgen}\\
  HDL& 2013 & High-Density-Lipid cholesterol & 94272& 992986 & \url{http://csg.sph.umich.edu//abecasis/public/lipids2013/} \\
  Ht & 2014 & Height & 252778& 827344 & \url{http://portals.broadinstitute.org/collaboration/giant/index.php/GIANT_consortium_data_files} \\
  LDL& 2013 & Low-Density-Lipid cholesterol & 89851& 990583 & \url{http://csg.sph.umich.edu//abecasis/public/lipids2013/} \\
  TC & 2013 & Total Cholesterol & 94556& 992889 & \url{http://csg.sph.umich.edu//abecasis/public/lipids2013/} \\
  TG & 2013 & Triglycerides & 90974& 990915& \url{http://csg.sph.umich.edu//abecasis/public/lipids2013/} \\
  RA & 2010 & Rheumatoid Arthritis & 25708 & 989551 & \url{http://www.broadinstitute.org/ftp/pub/rheumatoid_arthritis/Stahl_etal_2010NG/} \\
  T2D & 2008 & Type 2 Diabetes &63390& 1061515 & \url{http://diagram-consortium.org/downloads.html}\\
\hline
\end{tabular}
\end{table}
\end{landscape}


\section{Discussion}

In this study, we proposed a novel approach for high-dimensional regression analysis
when only marginal regression information and an external reference panel data set
are available.
Our work is motivated from combining information from multiple GWAS.
To date, a large number of GWAS have been conducted to find genetic factors associated with complex traits. Due to the need for privacy protection and issues in data-sharing of individual-level data, it is important to be able to effectively make full use of the summary statistics from separate studies.
In contrast to the limited sample size in individual-level data based GWAS analysis, a prominent feature of summary-level data analysis is that it can effectively make use of multiple data sets, which leads to a much larger combined sample size.

Under mild conditions, we prove that
the REMI estimator \eqref{remi_1} based on the marginal information and the reference penal  achieves the minimax optimal rate estimation error under reasonable conditions. In particular, the requirement on
the size of the reference panel data is quite mild, it is only in the order of the logarithm of the model
dimension.
Our theoretical result successfully explains why a relatively small reference sample
can be good enough for accurate estimation and prediction in real applications. We have conducted comprehensive simulations and real data analysis to demonstrate the utility of REMI.
The experimental results show that the performance of REMI can be very similar to the Lasso when the sample sizes of summary-level data and individual-level data are the same.
In genetic analysis,  summary-level data sets are much easier to access and their sample sizes are often orders of magnitude
larger than that of individual-level data sets. Consequently, REMI can be superior to the existing methods
requiring complete data  by taking advantages of the larger sample sizes, as demonstrated in our real data example.

\section*{Acknowledgment}

This work was supported in part by grants No. 11501579 and NO. 61501389
from National Science Funding of China, grants NO. 22302815, NO.
12316116 and NO. 12301417 from the Hong Kong Research Grant Council, and Initiation Grant NO. IGN17SC02 from University Grants Committee, startup grant R9405 from The Hong Kong University of Science and Technology, Shenzhen Fundamental Research Fund under Grant No. QTD2015033114415450 and grant R-913-200-098-263 from Duke-NUS Graduate Medical School, and AcRF Tier 2 (MOE2016-T2-2-029) from Ministry of Education, Singapore.

\section*{Appendix}

We recall some simple properties of subgaussian and subexponential
random variables.

\begin{lemma}\label{basic1}( Lemma 2.7.7 of \cite{Vershynin:2016} and  Remark 5.18 of  \cite{Vershynin:2010}.)
Let $\xi_1,\xi_2$ be sub-Gaussian random variables with noise level $\|\xi_1\|_{\psi_2} \leq \sigma_{{_{\xi_1}}}$ and $\|\xi_2\|_{\psi_2}\leq \sigma_{{_{\xi_2}}}$, respectively.   Then both $\xi_1\xi_2$ and $ \xi_1\xi_2- \mathbb{E}[\xi_1\xi_2]$ are sub-exponential random variables, and there exist an absolute constant $C > 0$ such that  $ \|\xi_1\xi_2- \mathbb{E}[\xi_1\xi_2]\|_{\psi_1} \leq C \sigma_{{_{\xi_1}}}\sigma_{{_{\xi_2}}}.$
\end{lemma}

We state the Bernstein-type inequality for the sum of independent and mean 0 sub-exponential
random random variables

\begin{lemma}\label{vershynin5.17}(Corollary  5.17 of \cite{Vershynin:2010})
Let $\xi_1,...,\xi_m$ be independent centered sub-exponential random variables.   Then for every $t>0$  one has $$\mathbb{P}[|\sum_{i = 1}^m \xi_i|/m \geq t] \leq  2 \exp (- C\min\{\frac{t^2}{K^2},\frac{t}{K}\}m),$$ where, $C$ is a absolute constant and $K = \max_{i = 1,..m}\{\|\xi_i\|_{\psi_1}\}$.
\end{lemma}

\begin{lemma}\label{linf}

Suppose the rows of $\bfX$ and $\bfX_{\mathrm{r}}$ are i.i.d sub-Gaussian samples drawn from  population with mean $\textbf{0}$ and covariance matrix $\bfSigma$.  Then, with probability at least $1-1/p^2$, we have  $$\|\widehat{\bfSigma} - \bfSigma \|_{\infty} \leq \frac{2C_1}{\sqrt{C}}\sqrt{\frac{\log p}{n}},$$
and $$\|\widehat{\bfSigma}_{\mathrm{r}}- \bfSigma\|_{\infty} \leq \frac{2C_1}{\sqrt{C}}\sqrt{\frac{\log p}{n_{\mathrm{r}}}},$$
as long as $n> \frac{4}{C}\log p$ and $n_{{\mathrm{r}}}> \frac{4}{C}\log p$.
\end{lemma}
\textbf{Proof of Lemma \ref{linf}.}
Since the proof of these two results are similar, we give one of them.
Let $\bfx_{i}$ be the $i-$th row of $\bfX$, $i =  1,...n$, and $(\bfx_{i})_{j}$ denote the $j-$th entry of $\bfx_{i}$.
Define $ G_{j,k}^{i} := (\bfx_{i})_{j}(\bfx_{i})_{k} - \mathbb{E}[(\bfx_{i})_{j}(\bfx_{i})_{k}]\in \mathcal{R}^{1}, i = 1,...,n, j= 1,...,p, k=1,..,p,$ which is sub-exponential with $\|G_{j,k}^{i}\|_{\psi_1} \leq C_1$ by Lemma \ref{basic1}.
Therefore,
\begin{align}
\mathbb{P}[\|\widehat{\bfSigma} - \bfSigma \|_{\infty}\geq t] &= \mathbb{P}[\|\sum_{i=1}^{n} (\bfx_i^{T}\bfx_i - \mathbb{E}[\bfx_i^T\bfx_i])/n \|_{\infty}\geq t]\nonumber\\
&= \mathbb{P}[\bigcup_{j= 1,k=1}^{p,p}|\sum_{i = 1}^{n} G_{j,k}^{i}/n| \geq t ]\nonumber\\
&\leq \sum_{j= 1, k=1}^{p,p} \mathbb{P}[|\sum_{i = 1}^{n} G_{j,k}^{i}|/n \geq t] \nonumber\\
&\leq p^2 \exp (- C\min\{\frac{t^2}{C_{1}^2},\frac{t}{C_1}\}n) \label{eq1}\\
& \leq p^2 \exp (- C \frac{t^2}{C_{1}^2}n)\nonumber
\end{align}
where the first inequality is due to union bound, and the second one follows from   Lemma \ref{vershynin5.17} and the last inequality  is because of
restricting $t\leq {C_1}$.
Then  Lemma \ref{linf} follows from  setting  $t = \frac{2C_1}{\sqrt{C}}\sqrt{\frac{\log p}{ n}}$ and the  assumption that $n> \frac{4}{C}\log p$.
$\hfill\Box$

The  following Lemma \ref{linfdif} - Lemma \ref{wnoise}   are building blocks for proving Theorem \ref{esterr}.

\begin{lemma}\label{linfdif}
Under the same assumption as Lemma \ref{linf}, we have  $$\|(\widehat{\bfSigma} - \widehat{\bfSigma}_{\mathrm{r}})\bfbeta^*_{\mathcal{A}} \|_{\infty} \leq \frac{2C_1C_3}{\sqrt{C}}(\sqrt{\frac{\log p}{n}} +\sqrt{\frac{\log p}{n_{\mathrm{r}}}}),$$ holds
with probability at least $1-2/p^2$.
\end{lemma}
\textbf{Proof of Lemma \ref{linfdif}.}
\begin{align*}
\|(\widehat{\bfSigma} - \widehat{\bfSigma}_{\textrm{r}})\bfbeta^*_{\mathcal{A}} \|_{\infty} &\leq  \|(\widehat{\bfSigma} - \bfSigma)\bfbeta^*_{\mathcal{A}}\|_{\infty} + \|(\bfSigma - \widehat{\bfSigma}_{\textrm{r}})\bfbeta^*_{\mathcal{A}}\|_{\infty}\\
&\leq  \|\widehat{\bfSigma} - \bfSigma\|_{\infty}\|\bfbeta^*_{\mathcal{A}}\|_{1} + \|\bfSigma - \widehat{\bfSigma}_{\textrm{r}}\|_{\infty}\|\bfbeta^*_{\mathcal{A}}\|_{1}\\
&\leq  \frac{2C_1C_3}{\sqrt{C}}\sqrt{\frac{\log p}{n}} + \frac{2C_1C_3}{\sqrt{C}} \sqrt{\frac{\log p}{{n_{\textrm{r}}}}},
\end{align*}
where the first inequality is due to triangle inequality, and the second inequality  follows from  Cauchy-Schwartz inequality, and the last   one holds with  probability larger than $1-2/p^2$ due to Lemma \ref{linf}. This finishes the proof of Lemma \ref{linfdif}.
$\hfill\Box$

\begin{lemma}\label{noise}
Suppose the rows of $\bfX$  are i.i.d sub-Gaussian samples drawn  from population with mean $\textbf{0}$ and covariance matrix $\bfSigma$, and the entries of  noise $\bfepsilon$ are i.i.d  centered sub-Gaussian with noise level $\sigma_{\epsilon}$.
With probability at least $1-1/p^3$, we have  $$\| \widetilde{\bfepsilon}\|_{\infty}  < 2\sigma_{\epsilon} \frac{C_1}{\sqrt{C}}\sqrt{\frac{\log p}{ n}},$$
provided that $n\geq \frac{4\log p}{C}.$
\end{lemma}
\textbf{Proof of Lemma \ref{noise}.}
We have,
\begin{align}
& \mathbb{P}[\|\widetilde{\bfepsilon}\|_{\infty}< t]=\mathbb{P}[ \|\bfX^{T}\bfepsilon/n\|_{\infty}< t]\nonumber\\
& = 1- \mathbb{P}[ \|\bfX^{T}\bfepsilon/n\|_{\infty}\geq t]\nonumber\\
&= 1- \mathbb{P}[\bigcup_{j= 1}^{p}|\bfX_j^T\bfepsilon/n | \geq t]\nonumber\\
&\geq 1- \sum_{ j=1}^{p} \mathbb{P}[|\sum_{i = 1}^{n} (\bfX_j)_{i} \epsilon_i|/n \geq t] \nonumber\\
&\geq 1- p \exp (- C\min\{\frac{t^2}{C_{1}^2\sigma^2_{\epsilon}},\frac{t}{C_1\sigma_{\epsilon}}\}n)\nonumber\\
& \geq1-  p \exp (- C\frac{t^2}{C_{1}^2\sigma^2_{\epsilon}}n)\nonumber\\
& \geq 1-1/p^3,\label{eq2}
\end{align}
the first  inequality is due to union bound, and the second  one follows from Lemma \ref{basic1} and   Lemma \ref{vershynin5.17}, where we use  $\|(\bfX_j)_i \epsilon_i -\mathbb{E}[(\bfX_j)_i \epsilon_i]\|_{\psi_1} \leq C \sigma_{\epsilon} C_1$,  and the last two  inequality  follows from
by setting $t = 2\sigma_{\epsilon} C_1\sqrt{\frac{\log p}{C n}}$ and the  assumption that $n> \frac{4}{C}\log p$, i.e., with probability at least
$1-1/p^3,$  we have $$ \|\widetilde{\bfepsilon}\|_{\infty}\leq 2\sigma_{\epsilon} C_1\sqrt{\frac{\log p}{C n}}.$$
$\hfill\Box$

\begin{lemma}\label{wnoise}
Under the same assumption as Lemma \ref{noise}, we have,
$$\|\widehat{\bfSigma} \bfbeta^*_{\mathcal{I}}\|_{\infty}\leq  \frac{2C_1C_4}{\sqrt{C}}\sqrt{\frac{\log p}{n}}  + 2C_2 \sigma_{\epsilon} \sqrt{\frac{\log p}{n}}.$$
\end{lemma}
with probability larger than $1-1/p^2$. 

\textbf{Proof of Lemma \ref{wnoise}.}
\begin{align*}
&\|\widehat{\bfSigma} \bfbeta^*_{\mathcal{I}}\|_{\infty} \leq  \|\bfSigma \bfbeta^*_{\mathcal{I}}\|_{\infty} +\|(\widehat{\bfSigma} -\bfSigma) \bfbeta^*_{\mathcal{I}}\|_{\infty} \\
&\leq   \max_{j=1,...p}\{\|\bfSigma_j\|_1\} \|\bfbeta^*_{\mathcal{I}} \|_{\infty} +\|\widehat{\bfSigma} -\bfSigma\|_{\infty}\|\bfbeta^*_{\mathcal{I}} \|_{1} \\
& \leq    2C_2 \sigma_{\epsilon} \sqrt{\frac{\log p}{n}} + \frac{2C_1C_4}{\sqrt{C}}\sqrt{\frac{\log p}{n}},
\end{align*}
where first inequality  is due to triangle inequality, and the second  one  follows from Cauchy-Schwartz inequality, the third inequality holds with probability larger than $1-1/p^2$ by using  \eqref{weaksparse} and  Lemma \ref{linf}.
This completes the proof of Lemma \ref{wnoise}.
$\hfill\Box$

Now we are ready to prove Theorem \ref{esterr}.

\textbf{Proof of Theorem \ref{esterr}.}
(i).
Let $\Delta = \widehat{\bfbeta}^{\textrm{c}} - \bfbeta^*_{\mathcal{A}}$.
Define the event $$\mathcal{E} = \{  2\|(\widehat{\bfSigma} - \widehat{\bfSigma_{r}}) \bfbeta^*_{\mathcal{A}}+ \widehat{\bfSigma} \bfbeta^*_{\mathcal{I}} + \widetilde{\bfepsilon}\|_{\infty}\leq \lambda_0 \}.$$
The optimality of $\widehat{\bfbeta}^{\textrm{c}}$ implies that
\begin{align}
 \langle \widehat{\bfbeta}^{\textrm{c}} , \widehat{\bfSigma}_{\textrm{r}} \widehat{\bfbeta}^{\textrm{c}} \rangle -{2}\langle\widetilde{\bfy},\widehat{\bfbeta}^{\textrm{c}} \rangle + \lambda \|\widehat{\bfbeta}^{\textrm{c}} \|_{1}&\leq \langle\bfbeta^*_{\mathcal{A}},\widehat{\bfbeta}^{\textrm{c}}\bfbeta^*_{\mathcal{A}} \rangle -{2}\langle\widetilde{\bfy},\bfbeta^*_{\mathcal{A}} \rangle + \lambda \|\bfbeta^*_{\mathcal{A}}\|_{1},\nonumber\\
&\Downarrow (\textrm{eq1})\nonumber\\ \langle\widehat{\bfbeta}^{\textrm{c}}-\bfbeta^*_{\mathcal{A}}, \widehat{\bfSigma}_{\textrm{r}}(\widehat{\bfbeta}^{\textrm{c}}-\bfbeta^*_{\mathcal{A}}) \rangle  + 2\langle\bfbeta^*_{\mathcal{A}},\widehat{\bfSigma}_{\textrm{r}}(\widehat{\bfbeta}^{\textrm{c}}-\bfbeta^*_{\mathcal{A}}) \rangle&+\lambda(\|\widehat{\bfbeta}^{\textrm{c}}{_\mathcal{A}}\|_1 + \|\widehat{\bfbeta}^{\textrm{c}}{_\mathcal{I}}\|_1)\leq
{2}\langle\widetilde{\bfy},\widehat{\bfbeta}^{\textrm{c}} - \bfbeta^*_{\mathcal{A}}  \rangle
 +\lambda\|\bfbeta^*_{\mathcal{A}}\|,\nonumber\\
& \Downarrow (\textrm{eq2})\nonumber\\
\langle\Delta,\widehat{\bfSigma}_{\textrm{r}}\Delta \rangle   +\lambda\|\Delta_{\mathcal{I}}\|_1\leq
{2}\langle\widetilde{\bfy},\Delta  \rangle & - 2\langle\widehat{\bfSigma}_{\textrm{r}}\bfbeta^*_{\mathcal{A}}, \Delta \rangle
 +\lambda\|\bfbeta^*_{\mathcal{A}}\| - \lambda\|\widehat{\bfbeta}^{\textrm{c}}{_\mathcal{A}}\|_1,\nonumber\\
& \Downarrow (\textrm{eq3})\nonumber\\
\langle\Delta,\widehat{\bfSigma}_{\textrm{r}}\Delta \rangle   +\lambda\|\Delta_{\mathcal{I}}\|_1  \leq
{2}\langle (\widehat{\bfSigma} & - \widehat{\bfSigma}_{\textrm{r}}) \bfbeta^*_{\mathcal{A}}+ \widehat{\bfSigma} \bfbeta^*_{\mathcal{I}}+ \widetilde{\bfepsilon},\Delta  \rangle
 +\lambda\|\Delta_{\mathcal{A}}\|,\nonumber\\
& \Downarrow (\textrm{eq4})\nonumber\\
\langle\Delta,\widehat{\bfSigma}_{\textrm{r}}\Delta \rangle   +\lambda\|\Delta_{\mathcal{I}}\|_1  \leq
{2}\|(\widehat{\bfSigma} &- \widehat{\bfSigma}_{\textrm{r}}) \bfbeta^*_{\mathcal{A}}+ \widehat{\bfSigma} \bfbeta^*_{\mathcal{I}}+ \widetilde{\bfepsilon}\|_{\infty}\|\Delta\|_1
 +\lambda\|\Delta_{\mathcal{A}}\|,\nonumber\\
 & \Downarrow (\textrm{eq5})\nonumber\\
 \langle\Delta,\widehat{\bfSigma}_{\textrm{r}}\Delta \rangle   +\lambda\|\Delta_{\mathcal{I}}\|_1  &\leq
\frac{\lambda}{2}(\|\Delta_{\mathcal{A}}\|_1  + \|\Delta_{\mathcal{I}}\|_1)
 +\lambda\|\Delta_{\mathcal{A}}\|,\nonumber\\
  & \Downarrow (\textrm{eq6})\nonumber\\
\langle\Delta,\widehat{\bfSigma}_{\textrm{r}}\Delta \rangle +\frac{\lambda}{2}\|\Delta_{\mathcal{I}}\|_1&\leq
\frac{3}{2} \lambda\|\Delta_{\mathcal{A}}\|_1,\label{basic2}
\end{align}
where, (eq1) and (eq2) and (eq6)  are due to some algebra, and (eq3) follows from \eqref{dec}, and (eq4) uses    Cauchy-Schwartz inequality, and (eq5) holds  by conditioning $\mathcal{E}$ and the assumption $\lambda_0\leq \lambda/2.$
It follow from \eqref{basic2} that
\begin{equation}\label{concd}
\Delta \in \mathcal{C}_{\mathcal{A},3}.
\end{equation}
Then, by the restricted eigenvalue condition on $\widehat{\bfSigma}_{\textrm{r}}$ and \eqref{basic2} we deduce,
$$\phi_0 \|\Delta\|_2^2 \leq \langle\Delta,\widehat{\bfSigma}_{\textrm{r}}\Delta\rangle\leq  \frac{3}{2} \lambda\|\Delta_{\mathcal{A}}\|_1\leq \frac{3}{2}\sqrt{s}\lambda\|\Delta\|_2,$$ i.e.,
$$\|\Delta\|_ 2 \leq \frac{3}{2\phi_0} \sqrt{s}\lambda\leq \frac{6}{\phi_0}(\frac{C_1(C_3+C_4)}{\sqrt{C}}\sqrt{\frac{s\log p}{n}}  +  \frac{C_1+\sqrt{C}C_2}{\sqrt{C}}\sigma_{\epsilon}\sqrt{\frac{s\log p}{ n}} + \frac{C_1C_3}{\sqrt{C}} \sqrt{\frac{s\log p}{n_{\textrm{r}}}}).$$
The above induction is conditioning on $\mathcal{E}.$ We need give a lower bound on $\mathbb{P}[\mathcal{E}]$.
Indeed,
\begin{align*}
{2}\|(\widehat{\bfSigma} - \widehat{\bfSigma}_{\textrm{r}}) \bfbeta^*_{\mathcal{A}}+ \widehat{\bfSigma} \bfbeta^*_{\mathcal{I}}+ \tilde{\bfepsilon}\|_{\infty} &\leq {2}\|(\widehat{\bfSigma} - \widehat{\bfSigma}_{\textrm{r}}) \bfbeta^*_{\mathcal{A}}\|_{\infty} + {2}\|\widehat{\bfSigma} \bfbeta^*_{\mathcal{I}}\|_{\infty} + {2}\| \tilde{\bfepsilon}\|_{\infty}\\
\end{align*}
Then, it follows  Lemma \ref{linfdif} and Lemma \ref{noise} and  Lemma \ref{wnoise} that
$\mathbb{P}[\mathcal{E}] \geq 1- 3/p^2 - 1/p^3.$
This completes the first part of proof of Theorem \ref{esterr}.

(ii). Let $\widehat{\bfSigma}_{\mathrm{new}} = \bfX_{\mathrm{new}}^{T} \bfX_{\mathrm{new}}/n_{\mathrm{new}}.$
Then,
\begin{align*}
&\|\bfX_{\mathrm{new}} (\widehat{\bfbeta}^{\textrm{c}} - \bfbeta^*_{\mathcal{A}})\|_2^2/n_{\mathrm{new}} = \langle\widehat{\bfSigma}_{\mathrm{new}}(\widehat{\bfbeta}^{\textrm{c}} - \bfbeta^*_{\mathcal{A}}), \widehat{\bfbeta}^{\textrm{c}} - \bfbeta^*_{\mathcal{A}}\rangle \\
&=  \langle\Delta, \widehat{\bfSigma}_{\mathrm{r}} \Delta \rangle +  \langle\Delta, (\widehat{\bfSigma}_{\mathrm{new}} - \widehat{\bfSigma}_{\mathrm{r}}) \Delta \rangle\\
& \leq \frac{3}{2}\lambda \|\Delta_{\mathcal{A}}\|_1 + \|\Delta\|_1^2 \|\widehat{\bfSigma}_{\mathrm{new}} - \widehat{\bfSigma}_{\mathrm{r}}\|_{\infty}\\
&\leq \frac{3}{2}\lambda \|\Delta_{\mathcal{A}}\|_1 +  \|\Delta\|_1^2 \frac{2C_1}{\sqrt{C}}(\sqrt{\frac{\log p}{n_{\mathrm{r}}}}+ \sqrt{\frac{\log p}{n_{\mathrm{new}}}})\\
& \leq  \mathcal{O}(\sigma_{\epsilon}\sqrt{\frac{s\log{p}}{n}} + \sqrt{\frac{s\log{p}}{n_{\mathrm{r}}}}) \|\Delta_{\mathcal{A}}\|_2  + \mathcal{O}(\sqrt{\frac{\log p}{n_{\mathrm{r}}}}+ \sqrt{\frac{\log p}{n_{\mathrm{new}}}}) s^2 \|\Delta_{\mathcal{A}}\|_2^2\\
&\leq  \mathcal{O}((\sigma_{\epsilon}\sqrt{\frac{s\log{p}}{n}} + \sqrt{\frac{s\log{p}}{n_{\mathrm{r}}}})^2) (1+ s^2 (\sqrt{\frac{\log p}{n_{\mathrm{r}}}}+ \sqrt{\frac{\log p}{n_{\mathrm{new}}}}) )\\
\end{align*}
where, the first inequality uses \eqref{basic2} and Cauchy-Schwartz inequality, and the second one is due to Lemma \ref{linf}, and the third inequality follow from \eqref{concd} and Cauchy-Schwartz inequality, the fourth inequality uses Theorem \ref{esterr}.
This completes the second part of proof of Theorem \ref{esterr}.
$\hfill\Box$

\section*{References}
\bibliographystyle{abbrv}

\end{document}